\documentclass{ws-mpla}
\usepackage{cite,slashed}
\usepackage{graphicx,graphbox,xcolor}
\usepackage{hyperref}

\renewcommand{\d}{\mathrm{d}}

\begin{document}

\markboth{A. Ghezal, Y. Delenda and M. Aouachria}{Constraining non-commutative geometry with W/Z+jet production at the LHC}

%%%%%%%%%%%%%%%%%%%%% Publisher's Area please ignore %%%%%%%%%%%%%%
\catchline{}{}{}{}{}
%%%%%%%%%%%%%%%%%%%%%%%%%%%%%%%%%%%%%%%%%%%%%%%%%%%%%%%%%%%%%%%%%%%

\title{Constraining non-commutative geometry with {\it{W/Z}}+jet \\production at the LHC}

\author{Achwaq Ghezal$^{1,\dag}$, Yazid Delenda$^{1,\ddag*}$\footnotetext{$^*$Corresponding author.}, Mekki Aouachria$^{1,2,\P}$}
\address{$^1$Laboratoire de Physique des Rayonnements et de leurs Interactions avec la Mati\`{e}re,\\
D\'{e}partement de Physique, Facult\'{e} des Sciences de la Mati\`{e}re,\\
Universit\'{e} de Batna-1, Batna 05000, Algeria\\
$^2$Laboratoire de Physique Energétique Appliquée,\\
D\'{e}partement de Physique, Facult\'{e} des Sciences de la Mati\`{e}re,\\
Universit\'{e} de Batna-1, Batna 05000, Algeria\\
$^\dag$achwaq.ghezal@univ-batna.dz\\
$^\ddag$yazid.delenda@univ-batna.dz\\
$^\P$mekki.aouachria@univ-batna.dz}

\maketitle

%\pub{Received  \\Revised }{Accepted  \\Published }

\begin{abstract}
We present a comprehensive calculation of the squared matrix elements for all partonic channels contributing to $W^\pm/Z$+jet production at hadron colliders within the framework of the non-commutative Standard Model (NCSM), including leptonic decays $W\to e\nu$ and $Z\to {\mu^+\mu^-}$. Our computation incorporates both $\mathcal{O}(\Theta)$ corrections to the Standard Model vertices and additional interaction terms inherent to the NCSM. A key finding is that the production amplitudes receive first-order corrections at $\mathcal{O}(\Theta)$, a distinctive feature compared to many other processes where non-commutative effects enter only at $\mathcal{O}(\Theta^2)$. The leptonic decay widths, in contrast, are modified solely at $\mathcal{O}(\Theta^2)$. This $\mathcal{O}(\Theta)$ enhancement provides improved sensitivity to non-commutative geometry, allowing us to probe for and constrain the non-commutative energy scale in the multi-TeV range. We provide numerical predictions for angular (azimuthal and rapidity) distributions and the forward--backward asymmetry, and compare them to state-of-the-art Standard Model predictions at leading and next-to-leading order from the \texttt{MCFM} Monte Carlo program. Finally, we test the NCSM with experimental data by analyzing an unbinned, particle-level $Z$+jet dataset from the ATLAS experiment. From this data, we calculate the azimuthal spectrum and forward-backward asymmetry, which are then used to derive stringent lower bounds on the non-commutative scale $\Lambda$. {Our analysis accounts for Earth rotation effects by treating the non-commutative tensor as fixed in a celestial frame and deriving time-averaged observables in the rotating detector frame.}
\end{abstract}

%\ccode{PACS Nos.: 13.60.Hb}

\section{Introduction}	

Current and future collider experiments play a central role in advancing high-energy particle physics. They enable unprecedented precision in the measurement of diverse kinematic distributions and observables~\cite{Golling:2016gvc}, thereby enhancing sensitivity to potential signals of physics beyond the Standard Model (SM). This work aims to directly bridge theoretical predictions and experimental measurements in such colliders, focusing on new physics scenarios motivated by the fact that the SM, widely regarded as a low-energy effective theory, is unable to fully account for several phenomena from both theoretical and experimental perspectives.

At the Large Hadron Collider (LHC), processes involving the production of a $W$ or $Z$ boson in association with a jet are powerful probes of new physics~\cite{Tricoli:2020uxr}. These processes provide precision tests of perturbative QCD, help constrain parton distribution functions (PDFs) and their uncertainties, and are sensitive to higher-order QCD and electroweak effects~\cite{Boughezal:2015dva,Boughezal:2016isb,Blumenschein:2018gtm}. From a phenomenological perspective, vector-boson-plus-jet production also constitutes a major background in searches for new particles. These features make it an excellent testing ground for scenarios that extend the Standard Model. In this work, we explore one such extension based on spacetime non-commutativity, examining its impact on single-jet production in association with a $W$ or $Z$ boson in proton–proton collisions. Our analysis focuses on leading-order effects arising from a minimal non-commutative (NC) deformation parameter $\Theta^{\mu\nu}$.

The concept of incorporating NC geometry into particle physics has a long history{, dating back to early ideas  of Heisenberg and Pauli \cite{Heisenberg:1929xj} with the objective of reducing the infinities that appear in quantum field theory. It subsequently evolved through the works of Snyder~\cite{Snyder:1946qz} and others~\cite{Yang:1947ud,Doplicher:1994zv} into NC structures, and was later established as a rigorous mathematical framework by Connes~\cite{Connes:1996gi}. The approach was subsequently extended to quantum field theories via the Moyal-Weyl product and the Seiberg-Witten map~\cite{Seiberg:1999vs}.} 

{NC geometry} has been studied in a variety of phenomenological contexts~\cite{Das:2007dn,Selvaganapathy:2015nva,Alboteanu:2006hh,Rizzo:2002yr,Manohar:2014zca,Wang:2011ei}, with particular motivation from quantum string theory~\cite{Seiberg:1999vs,Blumenhagen:2014sba}. Early developments faced challenges due to the non-local nature of non-commutativity and its apparent violation of Lorentz invariance~\cite{Douglas:2001ba,Calmet:2001na, Carroll:2001ws}.  {More generally, NC theories belong to a broader class of Lorentz-violating frameworks, which are systematically described by the Standard Model Extension (SME)~\cite{Colladay:1998fq,Kostelecky:2003fs,Kostelecky:2008ts,Aghababaei:2017gdd}. In this context, collider searches for Lorentz violation have been actively pursued \cite{Castorina:2010sj}; for instance, the CMS collaboration has recently placed stringent limits on Lorentz-violating couplings in top quark pair production~\cite{CMS:2024rcv}.} 

{Despite these challenges,} it has been shown that the low-energy limit of certain open-string models can be effectively described by Yang–Mills theories formulated on NC spaces~\cite{Alboteanu:2006hh}. Such models are theoretically appealing, offering a natural gauge structure and the possibility of probing Planck-scale physics at accessible energies~\cite{Calmet:2001na}, and more generally within the reach of future high-energy colliders~\cite{Golling:2016gvc}. In these frameworks, conventional spacetime coordinates $\hat{x}$ are replaced by NC operators satisfying
\begin{equation}
\left[\hat{x}^\mu,\hat{x}^\nu\right]=i\,\Theta^{\mu\nu}=i\,\frac{c^{\mu\nu}}{\Lambda^2}\,,
\end{equation}
where the hat denotes non-commutativity, $c^{\mu\nu}$ is a constant antisymmetric tensor, and $\Lambda$ is the characteristic energy scale at which NC effects become significant, typically assumed to be in the TeV range \cite{OPAL:2003eoc,Calmet:2001na}.

To construct a NC version of the SM, the structure of quantum fields must be deformed in accordance with the NC parameter $\Theta$. This is accomplished by replacing ordinary products with the Moyal-Weyl $\star$-product \cite{Douglas:2001ba,Riad:2000vy, Jurco:2000fb,Seiberg:1999vs}, defined by
\begin{equation}
(f\star g)(x)=\left.\exp\left(\frac{1}{2}\,\Theta_{\mu\nu}\,\partial_x^\mu\,\partial_y^\nu\right)f(x)\,g(y)\right|_{y=x}\,.
\end{equation}
This formulation introduces strong non-locality, resulting in novel features such as ultraviolet/infrared mixing and potential challenges to unitarity~\cite{Minwalla:1999px,Gomis:2000zz}. For a comprehensive review of the theoretical and technical aspects of NC field theory, see Ref.~\cite{Hinchliffe:2002km}.

The fields in NC gauge theories are related to their commutative counterparts through the Seiberg-Witten (SW) map~\cite{Ulker:2007fm,Seiberg:1999vs}, which ensures gauge covariance and establishes a consistent correspondence between the two formulations.  To first order in $\Theta$, the SW maps are given by~\cite{Batebi:2014lua,Melic:2005am,Jurco:2001rq}
\begin{subequations}
\begin{align}
\hat{A}_{\mu}&=A_{\mu}-\frac{1}{4}\,\Theta^{\nu\rho}\left\lbrace A_\nu,\partial_\rho A_\mu+F_{\rho\mu}\right\rbrace+\mathcal{O}(\Theta^2)\,,\\
\hat{\psi}&=\psi-\frac{1}{2}\,\Theta^{\mu\nu}\,A_\mu\partial_\nu\psi+\frac{i}{8}\,\Theta^{\mu\nu}\left[A_\mu,A_\nu\right]\psi+\mathcal{O}(\Theta^2)\,,
\end{align}
\end{subequations}
where $F_{\rho\mu}$ is the field strength tensor, $\{\, , \,\}$ denotes the anti-commutator, and $A_\mu$ and $\psi$ are the gauge and fermion fields, respectively.

The introduction of SW maps is essential for formulating a consistent effective theory on NC spacetime. In the minimal NC Standard Model (NCSM), proposed by Calmet \textit{et al.}~\cite{Calmet:2001na}, the gauge sector is restricted via the trace representation. A non-minimal extension, developed by Melić \textit{et al.}~\cite{Melic:2005am}, relaxes this constraint, allowing for additional interaction vertices, including neutral gauge boson self-couplings~\cite{Melic:2005fm,Melic:2005am}. These extensions lead to new Feynman rules and permit processes and decays that are forbidden in the conventional SM \cite{Wang:2011ei}. When the NC parameter $\Theta^{\mu\nu}$ vanishes, the theory continuously reduces to its commutative counterpart. However, constant time–space non-commutativity ($\Theta^{0i}\neq0$) is known to cause perturbative unitarity problems~\cite{Calmet:2001na,Gomis:2000zz}. These issues can be avoided by restricting to purely space–space non-commutativity or by adopting special NC structures.

In this work, we investigate the phenomenology of $W/Z$+jet production at the LHC within the NCSM, focusing on the leptonic decays $W \to e\nu$ and $Z \to {\mu^+\mu^-}$. We compute the squared amplitudes for all relevant partonic channels, incorporating both NC corrections to SM vertices and novel interaction terms intrinsic to the NCSM theory. Crucially, we demonstrate that NC effects emerge at $\mathcal{O}(\Theta)$, unlike many other processes where corrections appear only at $\mathcal{O}(\Theta^2)$. This distinctive feature significantly enhances sensitivity to the NC scale, potentially enabling constraints in the TeV range.

{Earth rotation introduces a time dependence in the NC tensor relative to a laboratory detector, as the latter rotates with respect to a fixed celestial frame. In this work, we account for this effect following previous works in the literature \cite{Kamoshita:2002wq, Das:2011iq, Manohar:2014zca} by treating the space–space deformation  as fixed in a celestial frame and explicitly deriving its time-dependent components in the rotating laboratory frame. This allows us to compute time-averaged observables and to assess the impact of sidereal modulation on the angular distributions, ensuring that our constraints on the NC scale $\Lambda$ are robust against Earth rotation effects.}

A key novelty of our work lies in performing a direct comparison between theoretical predictions in the NCSM framework and real LHC data from the ATLAS experiment, providing an experimental handle to constrain the fundamental scale of non-commutativity. By analyzing unbinned particle-level $Z$+jet data from ATLAS \cite{ATLAS:2024xxl,atlas_collaboration_2024_11507450}, we extract the azimuthal distribution and forward-backward asymmetry to derive robust experimental bounds on the NC deformation parameter. We also compare our results with predictions from the Monte Carlo program \texttt{MCFM} \cite{Campbell:2019dru} to assess the significance of NC corrections relative to NLO corrections.

This framework offers a compelling and experimentally testable extension of the SM. While prior studies have extensively explored NC effects, our work focuses specifically on $V$+jet interactions, a cornerstone of LHC physics, to uncover new insights and identify distinctive signatures of non-commutativity.

{Prior phenomenological studies of the NCSM at hadron colliders have focused on processes such as Drell--Yan processes~\cite{Selvaganapathy:2016jrl}, vector-boson pair production~\cite{Alboteanu:2006hh}, top-quark pair production~\cite{Fisli:2020vzt}, and $W^+W^-$ production~\cite{Ohl:2010zf}. In particular, Ref.~\cite{Alboteanu:2006hh} presented a leading-order study of $pp\to Z\gamma$ within the Seiberg--Witten framework and established characteristic kinematic signatures of NCSM effects at the LHC. Building on these earlier works, the present analysis extends the phenomenology by (i) deriving explicit analytic squared amplitudes for all partonic channels contributing to $W/Z+\text{jet}$ production; (ii) performing a full phenomenological convolution with modern PDFs and realistic experimental phase-space cuts using a Monte Carlo reweighting procedure developed for this analysis; (iii) carrying out a direct comparison with ATLAS Run-2 $Z+\text{jet}$ data through extracted kinematic distributions; and
(iv) incorporating Earth-rotation effects in deriving bounds on the NC scale $\Lambda$. These features distinguish the present work from previous NCSM collider studies and enable direct phenomenological constraints from $Z+\text{jet}$ data.}

This paper is organized as follows. In Section 2, we present the leading-order calculation of the squared amplitudes for the partonic subprocesses contributing to $pp \to W/Z+$jet production within the NCSM framework, making use of the \texttt{FeynCalc} package \cite{Shtabovenko:2020gxv} to perform the Dirac trace algebra. Section 3 provides a detailed cross-section analysis based on \texttt{MadGraph~5} \cite{Alwall:2011uj} and \texttt{MadAnalysis~5} \cite{Conte:2012fm}, highlighting the most relevant kinematic distributions and discussing the potential phenomenological implications of NC geometry. In Section 4, we compare the NCSM predictions with the SM results at leading and next-to-leading order (LO and NLO)  obtained using the \texttt{MCFM} Monte Carlo program, emphasizing the relative importance of NLO corrections with respect to NC effects. Section 5 presents an experimental study of $Z+$jet observables using publicly available ATLAS particle-level data, which provide high-precision measurements suitable for constraining deviations from the SM. The extracted distributions and the forward-backward asymmetry are then employed to set bounds on the fundamental NC scale $\Lambda$. Finally, conclusions are summarized in Section 6.

\section{\emph{W/Z}+jet production within the NCSM}

\begin{figure}[ht]
\centerline{\includegraphics[width=.7\textwidth]{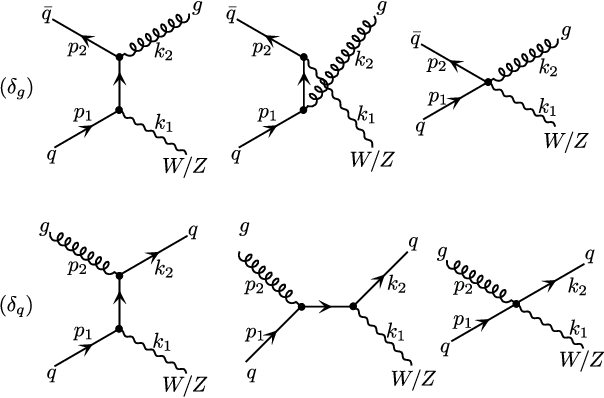}}
\vspace*{8pt}
\caption{Feynman diagrams for the leading-order contributions to $W/Z+$jet production in the NCSM at the LHC.\protect\label{fig1}}
\end{figure}

We study the process $pp\to W/Z+$jet at the LHC in the framework of the NCSM. To suppress contamination from large QCD backgrounds, we focus on leptonic decays of the vector bosons, specifically $W \to e\nu$ and $Z \to {\mu^+\mu^-}$. The incoming partons are treated as massless.  We adopt the standard shorthand notation for contractions involving the NC parameter
\begin{equation}
(\Theta\,k)^\mu=\Theta^{\mu\nu}\,k_\nu=-(k\,\Theta)^\mu\,, \quad (p\,\Theta\,k)=p_\mu\,\Theta^{\mu\nu}\,k_\nu\,.
\end{equation}
At leading order, two distinct partonic channels contribute to $W/Z+$jet production:
\begin{itemize}
\item $(\delta_g)$: $q\bar{q}' \to V + g$,
\item $(\delta_q)$: $q g \to V + q'$,
\end{itemize}
where $V$ denotes either a $W$ or $Z$ boson. The neutral-current process ($Z$ production) involves quarks of the same flavor, whereas the charged-current process ($W$ production) couples quarks of different flavors. Representative Feynman diagrams for these channels are shown in Fig.~\ref{fig1}, and the corresponding Feynman rules are listed in \ref{sec:Feyn}. A key feature of the NCSM is the appearance of a four-point gauge vertex, absent in the SM, which modifies the tree-level structure of these processes.

As shown in \ref{sec:decay}, NC geometry modifies the heavy-boson decay widths only at $\mathcal{O}(\Theta^2)$. In the narrow-width approximation, the production cross section factorizes from the decay branching ratio. Consequently, we compute NC corrections only at the production level, retaining terms up to $\mathcal{O}(\Theta)$.

The spin-, polarization-, and color-summed squared amplitudes for each channel are derived in detail in \ref{sec:Amps}. For a given channel $\delta$, the squared matrix element can be expressed as
\begin{equation}
|\mathcal{M}^{\delta}|^2=|\mathcal{M}^{\delta}_{\mathrm{SM}}|^2+|\mathcal{M}^{\delta}_{\mathrm{NC}}|^2+\mathcal{O}(\Theta^2)\,,
\end{equation}
where $|\mathcal{M}^{\delta}_{\mathrm{SM}}|^2$ is the SM result and $|\mathcal{M}^{\delta}_{\mathrm{NC}}|^2$ is the interference term between the SM and NC amplitudes, which is linear in $\Theta$.

For the gluon-initiated channel $(\delta_g)$, the results are
\begin{subequations}\label{eq:deltag}
\begin{align}
|\mathcal{M}^{\delta_g}_{\mathrm{SM}}|^2&=\mathcal{K}_V\,\frac{N^2-1}{N^2}\left(c_V^2+c_A^2\right)\frac{\hat{t}^2+\hat{u}^2+2\,\hat{s}\left(\hat{s}+\hat{t}+\hat{u}\right)}{\hat{t}\,\hat{u}}\,,\\
|\mathcal{M}^{\delta_g}_{\mathrm{NC}}|^2&=\mathcal{K}_V\,\frac{N^2-1}{N^2}\,2\,c_V\,c_A\,\frac{1}{\hat{t}\,\hat{u}}\,\Theta_{\mu\nu}\bigg[2\left(\hat{t}-\hat{u}\right)\epsilon^{\mu k_1p_1p_2}(p_1+p_2)^\nu\notag\\
&\quad+\frac{1}{2}\,\xi\left(\hat{s}+\hat{t}+\hat{u}\right)\left(\hat{t}\,\epsilon^{\mu\nu k_1k_2}+\left(\hat{t}-\hat{u}\right)\epsilon^{\mu\nu k_2p_2}\right)\bigg]\,,
\end{align}
\end{subequations}
and for the quark-initiated channel $(\delta_q)$
\begin{subequations}\label{eq:deltaq}
\begin{align}
|\mathcal{M}^{\delta_q}_{\mathrm{SM}}|^2&=-\mathcal{K}_V\,\frac{1}{N}\left(c_V^2+c_A^2\right)\frac{\hat{t}^2+\hat{s}^2+2\,\hat{u}\left(\hat{s}+\hat{t}+\hat{u}\right)}{\hat{t}\,\hat{s}}\,,\\
|\mathcal{M}^{\delta_q}_{\mathrm{NC}}|^2&=-\mathcal{K}_V\,\frac{1}{N}\,2\,c_V\,c_A\,\frac{1}{\hat{t}\,\hat{s}}\,\Theta_{\mu\nu}\bigg[2\left(\hat{t}-\hat{s}\right)\epsilon^{\mu k_1k_2p_1}(p_1-k_2)^\nu\notag\\
&\quad+\frac{1}{2}\,\xi\left(\hat{s}+\hat{t}+\hat{u}\right)\left(\hat{t}\,\epsilon^{\mu\nu p_2k_1}+\left(\hat{t}-\hat{s}\right)\epsilon^{\mu \nu p_2k_2}\right)\bigg]\,.
\end{align}
\end{subequations}
Here, $N=3$ is the number of QCD colors. The vector and axial-vector couplings are given by $c_V=T_3-2\,Q\sin^2\theta_w$ and $c_A=T_3$, where $T_3$ is the weak isospin and $Q$ is the quark electric charge. For up-type quarks ($u,c,t$), $T_3=+\frac{1}{2}$ and $Q=+\frac{2}{3}$; for down-type quarks ($d,s,b$), $T_3=-\frac{1}{2}$ and $Q=-\frac{1}{3}$. For $W+$jet production, we set $c_V=c_A=1$. The partonic Mandelstam variables are defined as
\begin{equation}
\hat{s}=(p_1+p_2)^2\,,\quad \hat{t}=(p_2-k_2)^2\,,\quad\hat{u}=(p_1-k_2)^2\,,
\end{equation}
and they satisfy $\hat{s}+\hat{t}+\hat{u}=m_V^2$. The antisymmetric tensor contractions are defined as
\begin{equation}
\epsilon^{\mu k_1p_1p_2}=\epsilon^{\mu\alpha\beta\delta}\,k_{1\alpha}\,p_{1\beta}\,p_{2\delta}\,,\quad \epsilon^{\mu \nu k_1 k_2}=\epsilon^{\mu\nu\beta\delta}\,k_{1\beta}\,k_{2\delta}\,,
\end{equation}
where $\epsilon^{\mu\alpha\beta\delta}$ is the Levi-Civita tensor ($\epsilon^{0123} = +1$). The bookkeeping parameter $\xi=1$ is used to track contributions from the four-point vertex diagram.

The overall coupling factors for the $W$ and $Z$ processes are, respectively,
\begin{subequations}
\begin{align}
\mathcal{K}_W&=\frac{e^2\,g_s^2}{8\sin^2\theta_w}\,|V_{ij}|^2\,,\\
\mathcal{K}_Z&=\frac{e^2\,g_s^2}{4\sin^2\theta_w\cos^2\theta_w}\,,
\end{align}
\end{subequations}
where $e$ is the electromagnetic coupling, $g_s$ is the strong coupling constant, $\theta_w$ is the Weinberg angle, and $V_{ij}$ is the relevant CKM matrix element for the $W$ boson interaction.

{Before proceeding to the phenomenological predictions, we briefly discuss the domain of validity of the $\mathcal{O}(\Theta)$ expansion. The amplitudes in Eqs.~\eqref{eq:deltag} and \eqref{eq:deltaq} are derived from the SW map truncated at first order in the NC parameter $\Theta^{\mu\nu}$. This truncation should be viewed as an effective expansion, expected to be reliable when the characteristic hard scale $Q$ of the process remains below the fundamental NC scale $\Lambda$. For $Q\gtrsim \Lambda$, higher-order terms in the SW expansion, $\mathcal O(\Theta^2)$ and beyond, may become important, potentially reducing the accuracy and perturbative control of the truncated $\mathcal O(\Theta)$ description.}

{Moreover, issues related to unitarity have been discussed in the literature, particularly for theories with time-like non-commutativity~\cite{Gomis:2000zz,Calmet:2001na}. For the purely space-space non-commutativity considered throughout this work, such pathologies are absent or substantially milder; nevertheless, the validity of the truncated expansion still motivates restricting the relevant hard scale. As a conservative consistency check, we therefore consider imposing the kinematic condition $Q<\Lambda$, implemented using the hard-scale choice $Q=p_t$. This excludes only the extreme high-$p_t$ tail where the $\mathcal O(\Theta)$ expansion could potentially become unreliable.} 

\section{\emph{W/Z}+jet phenomenology in NC geometry}

We now present a phenomenological study of $W/Z$+jet production at the LHC within the NCSM framework. Using the squared amplitudes derived up to $\mathcal{O}(\Theta)$ in the previous section, we evaluate a set of differential distributions that are directly relevant for LHC analyses.

The choice of observables depends on the specific vector boson. For the $W$+jet channel, the presence of an undetected neutrino (manifesting as missing transverse energy) prevents a full kinematic reconstruction of the $W$ boson. Consequently, we focus on differential observables built from the charged lepton (electron) in the final state, which can be measured with high precision. In contrast, the $Z$+jet process allows for a complete kinematic reconstruction of the $Z$ boson from its dilepton decay products, enabling a direct study of the $Z$-boson properties.

To circumvent potential unitarity issues associated with time-space or time-time non-commutativity~\cite{Gomis:2000zz}, we restrict our analysis to the case of purely space-space non-commutativity. This choice is theoretically well motivated and finds support in certain string-theoretic constructions. We parameterize the NC tensor $\Theta^{\mu\nu}$ in direct analogy to a magnetic field strength tensor
\begin{equation}
\Theta^{\mu\nu} =\frac{c^{\mu\nu}}{\Lambda^2}\,,\quad
c^{\mu\nu}=\begin{pmatrix}
0 & 0&0 & 0 \\
0 & 0 & -\beta_z&\beta_y\\
0 &\beta_z & 0& -\beta_x\\
0 & -\beta_y & \beta_x &0\\
\end{pmatrix},
\end{equation}
where $\Lambda$ denotes the fundamental energy scale of non-commutativity, and the dimensionless parameters $\beta_i$ ($i = x, y, z$) quantify the strength of non-commutativity along each spatial direction.

{At leading order in non-commutativity, the corrections to the squared amplitudes, and hence to all observables considered here, enter linearly in the deformation tensor, $|\mathcal M|^2 = |\mathcal M|^2_{\rm SM} + \Theta^{\mu\nu} C_{\mu\nu}$,
where $C_{\mu\nu}$ encodes the process-dependent interference between the SM and NC amplitudes. This linear structure implies that the independent components of $\Theta^{\mu\nu}$ may be analyzed separately, while a generic orientation can, at this order, be expressed in terms of a linear combination of these basis contributions. The study of the individual components of $\boldsymbol{\beta}$ therefore serves as a decomposition of the general NC geometry, rather than a choice of physically preferred benchmark orientations. In the following sections, we first investigate the phenomenological impact of each component of $\boldsymbol{\beta}$ on collider observables, and subsequently incorporate Earth-rotation effects to analyze fully general orientations relevant for comparison with ATLAS data.}

\subsection{Total cross-section calculation}

The total cross-section for $W/Z$+jet production is obtained by summing over all contributing partonic channels:
\begin{equation}
\sigma_{\mathrm{tot}}=\mathcal{B}\,\sum_{\delta}\int\d x_a\,\d x_b\,f_a(x_a,\mu_f^2)\,f_b(x_b,\mu_f^2)\,\Xi\,\frac{\d p_t^2\,\d y\,\d\phi}{32\pi^3 \hat{s}}\,\delta\left(\hat{s}+\hat{t}+\hat{u}-m_V^2\right)\left|\mathcal{M}^\delta\right|^2,
\end{equation}
where $\mathcal{B}$ denotes the branching ratio for the $W/Z$ decay, and $f_i(x_i,\mu_f^2)$ is the PDF for an incoming parton of flavor $i$ carrying a momentum fraction $x_i$ at the factorization scale $\mu_f$. The factor $\Xi$ implements the experimental selection cuts. The variables $p_t$, $y$, and $\phi$ represent the transverse momentum, rapidity, and azimuthal angle of the final-state jet, respectively, and $m_V$ is the vector boson mass.

We compute both total and differential cross-sections using a reweighting technique applied to a Monte-Carlo--generated SM event sample. Unweighted parton-level events for the SM processes $pp\to W(e\nu)/Z({\mu\mu})+$jet are generated with \texttt{MadGraph5} \cite{Maltoni:2002qb, Alwall:2014hca}. The events are then analyzed with \texttt{MadAnalysis5} \cite{Conte:2012fm} to assign modified weights according to
\begin{equation}
\text{weight}=\left(1+\frac{|\mathcal{M}_{\mathrm{NC}}|^2}{|\mathcal{M}_{\mathrm{SM}}|^2}\right)\sigma_{\mathrm{tot,SM}}\,,
\end{equation}
where $\sigma_{\mathrm{tot,SM}}$ is the SM total cross-section computed by \texttt{MadGraph5}. The reweighting is performed event-by-event using the partonic kinematics, and the total NCSM cross-section is obtained as the mean of all weights across the event sample.

This procedure is valid because the kinematic phase space of the NCSM is identical to that of the SM, with deviations arising only from modified interaction vertices. Thus, one can reliably obtain NCSM predictions by reweighting SM event samples with the appropriate ratio of squared matrix elements. 

Our analysis considers proton-proton collisions at a center-of-mass energy of $\sqrt{s}=13~\mathrm{TeV}$, using the \texttt{NNPDF40\_nlo\_as\_01180} PDF set~\cite{NNPDF:2021njg} interfaced through \texttt{LHAPDF6}~\cite{Buckley:2014ana}.\footnote{{Standard QCD scaling violations are implicitly accounted for through the DGLAP-evolved PDFs employed in our numerical analysis, evaluated at the factorization scale $\mu_f=m_V$. These effects are therefore included in the SM baseline used to assess the NC corrections.}} We apply the kinematic cuts $|y| < 5$ on the rapidity and $p_t > 50~\mathrm{GeV}$ on the transverse momentum of the final-state jet. The renormalization and factorization scales are set to the vector boson mass, $\mu_r = \mu_f = m_V$, with $m_W = 80.419~\mathrm{GeV}$ and $m_Z = 91.188~\mathrm{GeV}$.\footnote{Masses and couplings are taken from the default parameters of the \texttt{NNPDF40\_nlo\_as\_01180} set.}

Table \ref{tab1} summarizes the calculated total cross-sections for $pp\to W(e\nu)/Z({\mu\mu})+$jet production in the NCSM. The left panel shows the dependence on the NC scale $\Lambda$, while the right panel presents the effect of the NC orientation, parameterized by the unit vector $(\beta_x,\beta_y,\beta_z)$, for a fixed $\Lambda = 300~\mathrm{GeV}$.
\begin{table}[ht]
\tbl{\label{tab1} 
Leading order cross-sections (in pb) for $pp\to W(e\nu)/Z({\mu\mu})+$jet processes in the NCSM at $\sqrt{s}=13$ TeV. Left: dependence on the NC scale $\Lambda$. Right: dependence on NC orientation at $\Lambda=300$ GeV.}
{\begin{tabular}{@{}ccc@{}} \toprule
$\Lambda$ (GeV) & $W(e\nu)+$jet & $Z({\mu\mu})$+jet \\
\colrule
$\infty$     & 876.78 & 101.22\\
1000         & 876.76 & 101.23\\
700          & 876.73 & 101.23\\
500          & 876.68 & 101.24\\
300          & 876.51 & 101.26\\
 \botrule
\end{tabular}
\qquad\qquad\begin{tabular}{@{}ccc@{}} \toprule
$(\beta_x,\beta_y,\beta_z)$  & $W(e\nu)+$jet & $Z({\mu\mu})+$jet \\
\colrule
(1,0,0)  & 876.74 & 101.25 \\
(0,1,0)  & 876.58 & 101.23 \\
(0,0,1)  & 876.77 & 101.23 \\
 \botrule
\end{tabular}}
\end{table}

The results indicate that deviations from SM predictions are negligible, even for relatively low NC scales ($\Lambda=300~\mathrm{GeV}$). This demonstrates that the total inclusive cross-section is not a sensitive probe of space-space non-commutativity at current LHC energies, suggesting the need for more differential observables to uncover potential NC effects.

{As a validation test of the truncated $\mathcal O(\Theta)$ expansion, we examine the impact of imposing the conservative kinematic condition $p_t<\Lambda$ at the Monte Carlo event-generation stage, rejecting events with $p_t\ge \Lambda$. The results are shown in Table~\ref{tab1_1} for the process $pp\to Z+$jet. For the $\Lambda$ values relevant to our analysis, the overwhelming majority of events populate the region well below this validity cut, owing to the strong suppression of the high-$p_t$ tail by the PDFs. We find that the fraction of simulated events failing this condition is extremely small, below $0.35\%$ even for the lowest value of $\Lambda$ considered, and decreasing further for larger $\Lambda$. The resulting change in the total cross section is correspondingly negligible, indicating that our phenomenological conclusions and extracted limits are insensitive to this auxiliary consistency cut.}

\begin{table}[ht]
\tbl{\label{tab1_1} 
Impact of imposing the cut $p_t <\Lambda$ on the MC event sample and on the cross-section for the process $pp\to Z(\mu\mu)+$ jet.}
{\begin{tabular}{@{}ccc@{}} \toprule
$\Lambda$ (GeV) & Rejected events & Cross-section \\
\colrule
$\infty$     & --- & 101.22\\
1000         & 0.0006\% & 101.23\\
700          & 0.0047\% & 101.25\\
500          & 0.029\%  & 101.21\\
300          & 0.33\%   & 100.93\\
 \botrule
\end{tabular}}
\end{table}

{For this reason, and to maintain a common fiducial phase space in the comparison with ATLAS data, the experimental analysis and bounds presented in Sec.~6 are shown without imposing this auxiliary cut.}

\subsection{Angular distributions}

We now calculate differential distributions for key angular observables, which offer enhanced sensitivity to NC effects compared to total cross-section measurements. This sensitivity arises from their dependence on the relative orientation between the NC deformation axis and the final-state momentum directions.

For $W(e\nu)+$jet production, we examine the electron rapidity $\eta_e$ and azimuthal angle $\phi_e$ from the $W$ boson decay. For $Z({\mu\mu})+$jet events, we analyze the rapidity $\eta_Z$ and azimuthal angle $\phi_Z$ of the reconstructed $Z$ boson from its dilepton decay products. The rapidity is defined as
\begin{equation}
\eta = \frac{1}{2}\ln\frac{E+p_z}{E-p_z}\,.
\end{equation}

\begin{figure}[ht]
\centerline{\includegraphics[width=.49\textwidth]{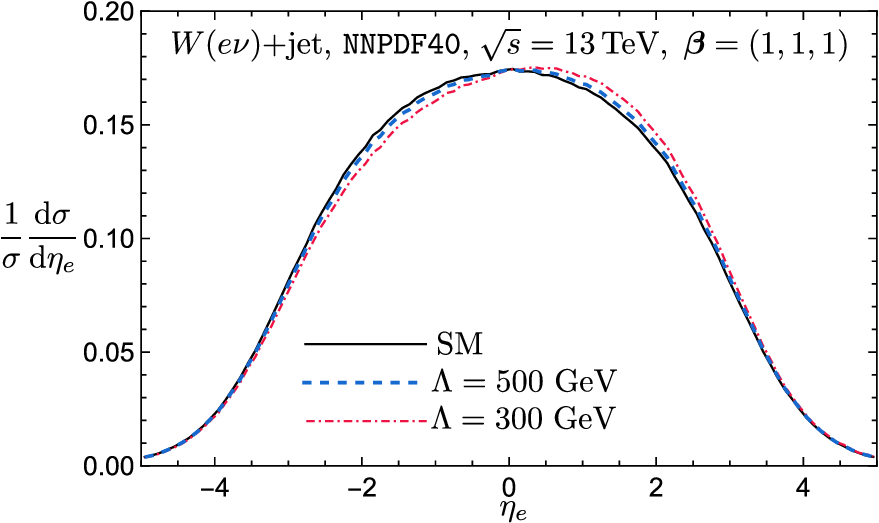}
\includegraphics[width=.49\textwidth]{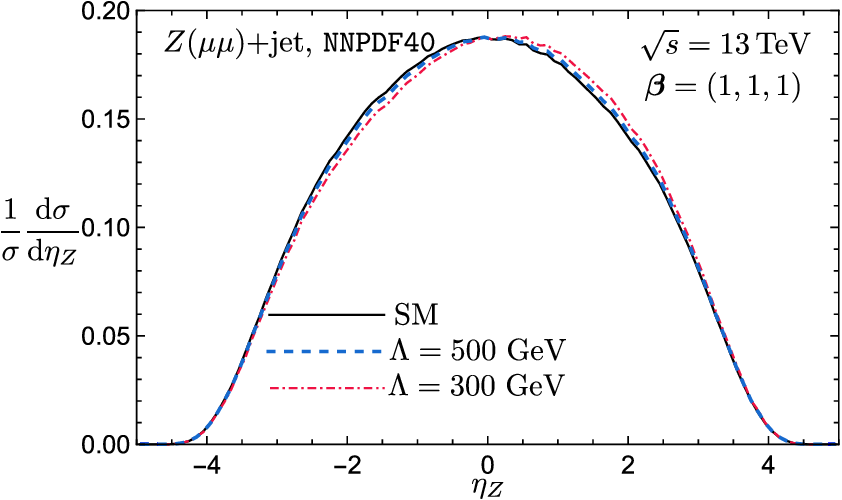}}
\centerline{
\includegraphics[width=.49\textwidth]{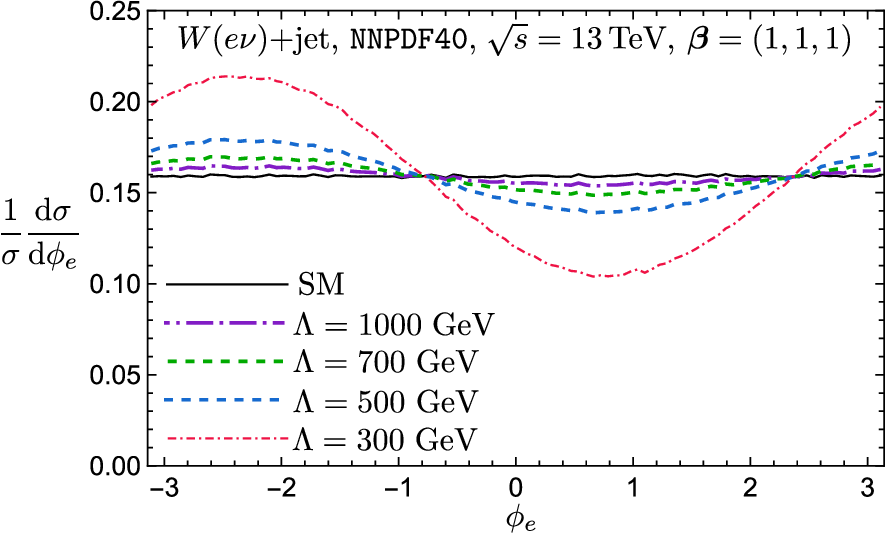}
\includegraphics[width=.49\textwidth]{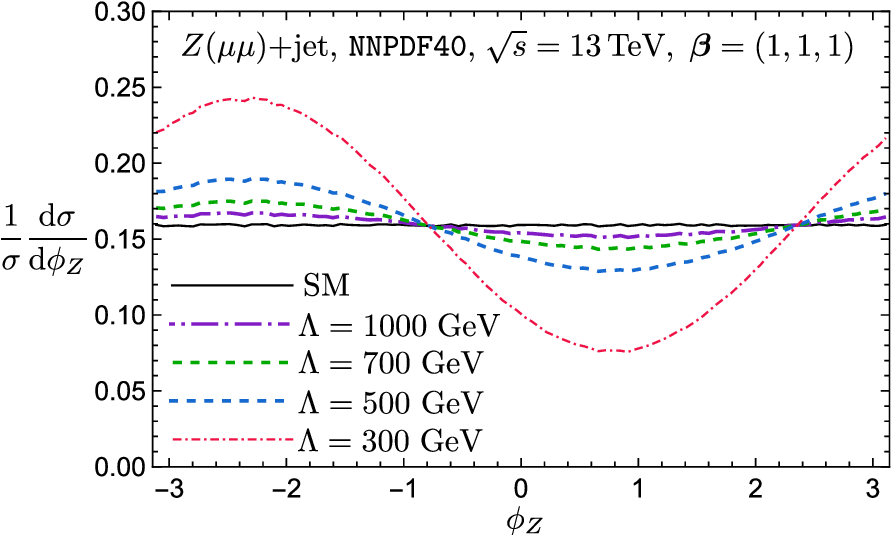}}
\vspace*{8pt}
\caption{Normalized differential rapidity (top) and azimuthal (bottom) distributions for various NC scales. Left column: electron from $W$ decay. Right column: $Z$ boson.\protect\label{fig2}}
\end{figure}

\begin{figure}[ht]
\centerline{\includegraphics[width=.49\textwidth]{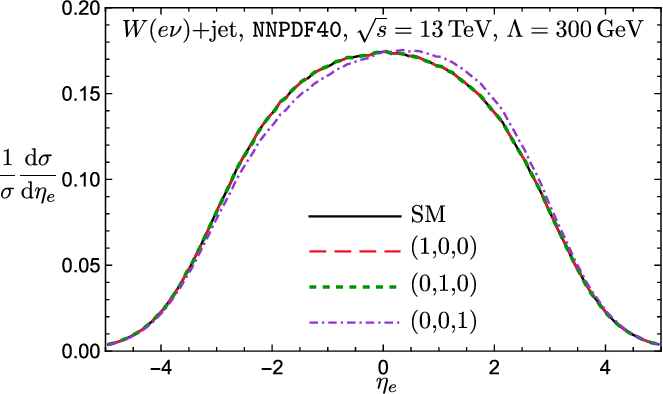}
\includegraphics[width=.49\textwidth]{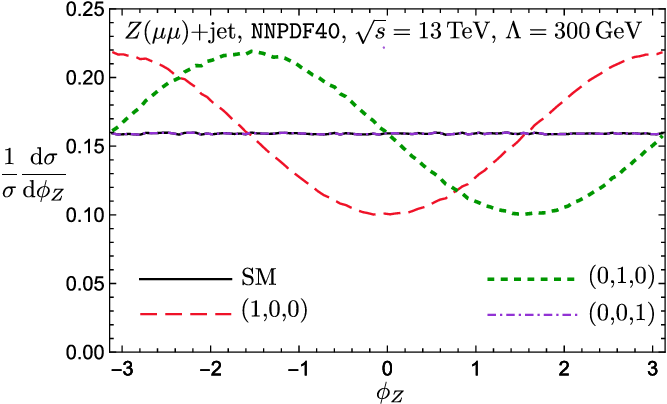}}
\vspace*{8pt}
\caption{
Normalized differential rapidity (left) and azimuthal (right) distributions for the three principal directions of the NC tensor, $(\beta_x, \beta_y, \beta_z) = (1,0,0)$, $(0,1,0)$, and $(0,0,1)$, at a fixed NC scale $\Lambda$.\protect\label{fig3}}
\end{figure}

The differential distributions are obtained using the reweighting method described in the previous subsection, applied to each bin of the histogram. Figure~\ref{fig2} displays the normalized differential distributions for these observables for various NC scales, while Figure~\ref{fig3} shows the corresponding distributions for the three basis directions of the NC tensor.

The SM prediction for the rapidity distribution exhibits perfect symmetry about $\eta=0$. This symmetry is broken in the NCSM for NC deformations along the beam direction, $(0,0,1)$. The resulting asymmetry is quantified by the forward-backward asymmetry
\begin{equation}
A_{\mathrm{FB}}=\frac{\sigma(\eta>0)-\sigma(\eta<0)}{\sigma(\eta>0)+\sigma(\eta<0)}\,.
\end{equation}
Values of $A_{\mathrm{FB}}$ for the electron from $W$ decay ($A_{\mathrm{FB}}^e$) and for the $Z$ boson ($A_{\mathrm{FB}}^Z$) are summarized in Table~\ref{tab:afb} for various NC scales.
\begin{table}[ht]
\tbl{Forward-backward asymmetry $A_{\mathrm{FB}}$ of the electron from $W$ decay ($A_{\mathrm{FB}}^e$) and of the $Z$ boson ($A_{\mathrm{FB}}^Z$) for various NC scales.}
{\begin{tabular}{@{}ccc@{}} \toprule
$\Lambda$ (GeV) & $A_{\mathrm{FB}}^e$ & $A_{\mathrm{FB}}^Z$\\
\colrule
SM       & 0.0000 & 0.0000 \\
1000 GeV & 0.0034 & 0.0028 \\
700 GeV  & 0.0069 & 0.0057 \\
500 GeV  & 0.0135 & 0.0111 \\
300 GeV  & 0.0375 & 0.0308 \\
 \botrule
\end{tabular}\label{tab:afb} }
\end{table}

For the azimuthal distribution, significant deviations from isotropy are observed for NC deformations along the transverse directions, $(1,0,0)$ and $(0,1,0)$, with the two patterns phase-shifted by $\pi/2$ relative to each other. These deviations are well described by a simple sinusoidal modulation
\begin{equation}\label{eq:sin}
\frac{1}{\sigma}\frac{\d\sigma}{\d\phi} = \frac{1}{2\pi}-\frac{\Lambda_0^2}{\Lambda^2}\left(\beta_x\cos\phi+\beta_y \sin \phi\right),
\end{equation}
where the scale $\Lambda_0$ is proportional to the jet's transverse momentum $p_t$. Different values of $\beta_x$ and $\beta_y$ induce a phase shift in the distribution with amplitude of oscillation proportional to $\sqrt{\beta_x^2+\beta_y^2}$.

The azimuthal distribution exhibits remarkable sensitivity to NC effects, providing constraints competitive with scales on the order of $1~\mathrm{TeV}$. Imposing a high-$p_t$ cut on jets further enhances this sensitivity, making it a powerful observable for probing NC-induced anisotropies in the plane transverse to the beam. Conversely, the rapidity distribution and its associated $A_{\mathrm{FB}}$ are more sensitive to NC effects along the beam axis.

\begin{figure}[ht]
\centerline{\includegraphics[width=.49\textwidth]{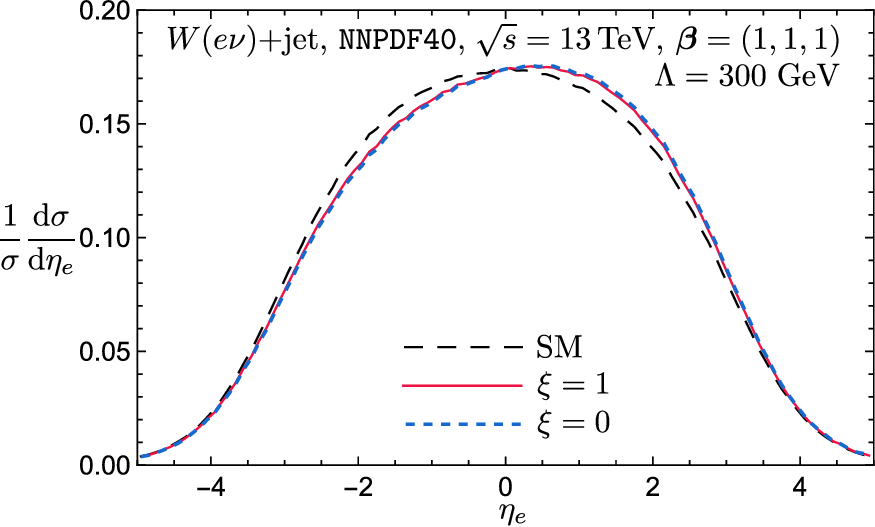}
\includegraphics[width=.49\textwidth]{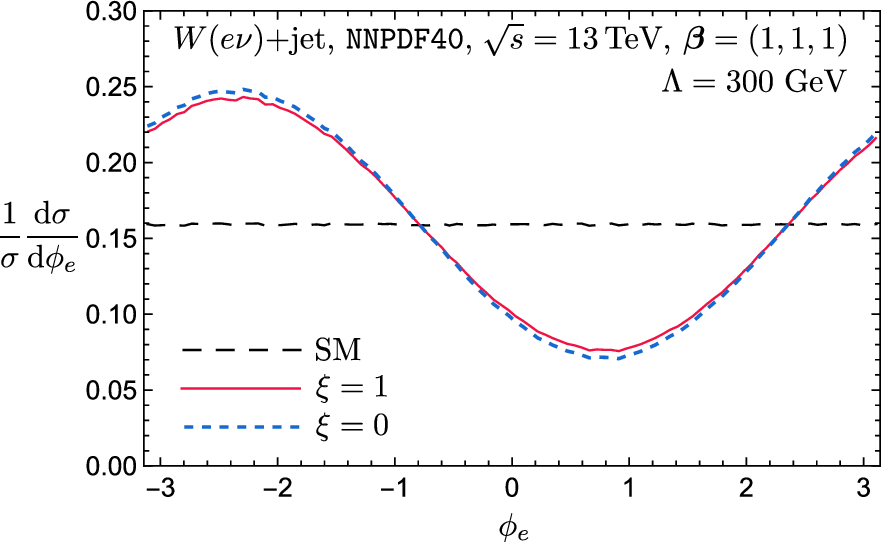}}
\vspace*{8pt}
\caption{Impact of the four-point interaction on the angular distributions. The parameter $\xi$ controls the inclusion ($\xi=1$) or exclusion ($\xi=0$) of the four-point vertex contribution. The left panel shows the rapidity distribution, the right panel the azimuthal distribution.\protect\label{fig4}}
\end{figure}

The observed modulation is dominated by contributions from the deformed SM-like vertices. The four-point vertex, unique to the NCSM, contributes only a minor distortion. As shown in Fig.~\ref{fig4}, switching off this interaction ($\xi = 0$) results in negligible changes to the distributions, confirming the subdominant role of this vertex at $\mathcal{O}(\Theta)$.

\section{Comparison to fixed-order Monte Carlo simulations}

\begin{figure}[ht]
\centerline{\includegraphics[width=.49\textwidth]{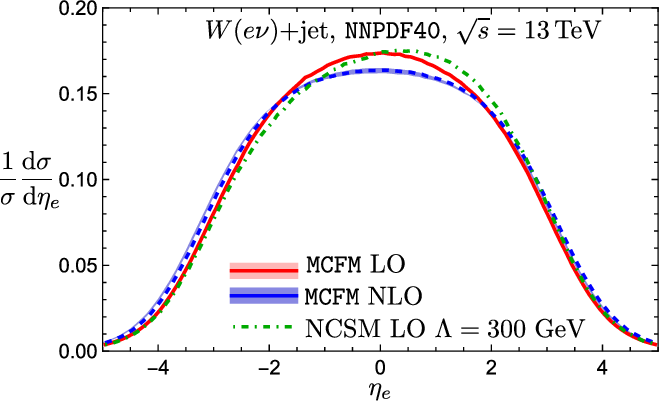}
\includegraphics[width=.49\textwidth]{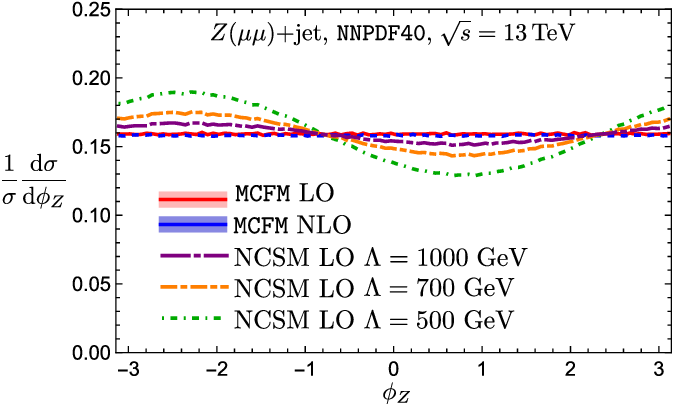}}
\vspace*{8pt}
\caption{
SM angular distributions obtained with \texttt{MCFM} at LO and NLO, compared to predictions from the NCSM. The bands represent theoretical uncertainties from renormalization and factorization scale variations.\protect\label{fig5}}
\end{figure}

To assess the relative importance of NC effects against the higher-order SM corrections, we compare our NCSM predictions with fixed-order SM predictions obtained using the \texttt{MCFM} program~\cite{Campbell:2019dru}. In particular, We compute the relevant differential distributions at both LO and NLO in QCD and contrast them with the NCSM predictions. The results of this comparison for the rapidity and azimuthal observables are shown in Fig.~\ref{fig5}.

Theoretical uncertainties for both the LO and NLO SM predictions are estimated through a standard six-point variation of the renormalization and factorization scales. As expected, these uncertainties are significantly reduced in the normalized distributions, affirming their suitability as precision observables for detecting deviations from the SM.

The comparison reveals two key results. For the rapidity distribution, a noticeable difference exists between the LO and NLO SM predictions: the NLO corrections broaden the distribution and lower the central peak. Crucially, however, the symmetry around $\eta=0$ is preserved to a high degree of accuracy. Consequently, the forward-backward asymmetry $A_{\mathrm{FB}}$ remains consistent with zero at both LO and NLO. This feature highlights the robustness $A_{\mathrm{FB}}$ as a clean probe of NC effects, as any statistically significant nonzero measurement could be unambiguously attributed to new physics. For the azimuthal distribution, the SM prediction remains perfectly flat at both LO and NLO. Furthermore, the NLO corrections and the associated scale variations have a negligible impact on the normalized distribution. This demonstrates a remarkable stability against higher-order QCD effects, making any deviation from flatness a clean signature of NC-induced anisotropy.

This comparison with fixed-order simulations demonstrates that while SM NLO corrections are quantitatively important for absolute cross-sections, they have a minimal impact on the \emph{normalized} angular observables central to our analysis. The stability of the SM prediction for these observables reinforces their selection as powerful and robust discriminants for NC effects, which we constrain with experimental data in the next section.

\section{{Earth rotation effects}}

Before proceeding to comparison with real-world experimental data, we need to examine how Earth's rotation affects the angular distributions. We assume that the NC reference frame is fixed with respect to the cosmic microwave background and neglect the motion of the Sun over the duration of the experiment. Under this approximation, only Earth's rotation about its axis is relevant. After one sidereal day, the detector orientation returns to its initial configuration.

Following Ref. \cite{Kamoshita:2002wq}, the fixed NC basis $(\boldsymbol{i}_\mathrm{NC},\boldsymbol{j}_\mathrm{NC},\boldsymbol{k}_\mathrm{NC})$ can be expressed in terms of the detector-frame basis $(\boldsymbol{i},\boldsymbol{j},\boldsymbol{k})$ as
\begin{subequations}
\begin{align}
\boldsymbol{i}_\mathrm{NC}&=\left(c_a s_\zeta + s_\delta s_a c_\zeta\right)\boldsymbol{i}+c_\delta c_\zeta \,\boldsymbol{j} + \left(s_a s_\zeta - s_\delta c_a c_\zeta\right) \boldsymbol{k}\,,\\
\boldsymbol{j}_\mathrm{NC}&= \left(-c_a c_\zeta + s_\delta s_a s_\zeta\right)\boldsymbol{i} + c_\delta s_\zeta \,\boldsymbol{j} + \left(-s_a c_\zeta - s_\delta c_a s_\zeta\right)\boldsymbol{k}\,,\\
\boldsymbol{k}_\mathrm{NC}&= -c_\delta s_a\, \boldsymbol{i} + s_\delta \,\boldsymbol{j} + c_\delta c_a \,\boldsymbol{k}\,,
\end{align}
\end{subequations}
where $\delta$ is the detector's geographical latitude, and $a$ denotes the azimuthal orientation of the $+z$ (beam) axis relative to geographic North (positive sense being counterclockwise toward the west). We define $\zeta=\omega t$, with $\omega=2\pi/T$ and $T=23.934$ hours (the sidereal day). The shorthand $c_\alpha$ and $s_\alpha$ denote $\cos\alpha$ and $\sin\alpha$, respectively. In this setup, Earth's rotation axis is aligned with $\boldsymbol{k}_\mathrm{NC}$.

\begin{figure}[ht]
\centerline{\includegraphics[width=.65\textwidth]{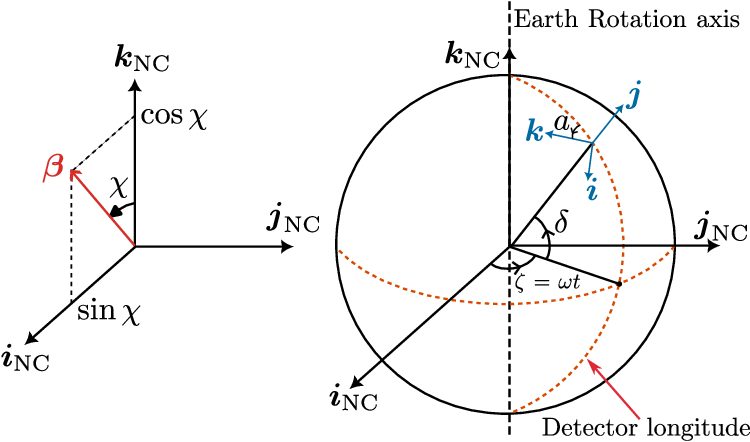}}
\vspace*{8pt}
\caption{Relation between the fixed NC coordinate system and the detector (laboratory) frame. The laboratory $y$ axis points vertically upward, while the $x$ and $z$ axes lie in the tangent plane.\protect\label{Earth}}
\end{figure}

Without loss of generality, we choose the NC space-space deformation vector $\boldsymbol{\beta}$ to lie in the $(\boldsymbol{i}_\mathrm{NC},\boldsymbol{k}_\mathrm{NC})$ plane,
\begin{equation}
\boldsymbol{\beta}= \cos\chi\,\boldsymbol{k}_\mathrm{NC} + \sin\chi\,\boldsymbol{i}_\mathrm{NC}\,,
\end{equation}
with $0\leq\chi\leq \pi$. The coordinate systems are illustrated in Fig.~\ref{Earth}.

Because the NC corrections to the squared amplitudes (and consequently to the distributions) are linear in $\Theta^{\mu\nu}$ (i.e., $\mathcal{O}(\Theta)$), they can be decomposed into contributions from the three components of $\boldsymbol{\beta}$. As shown in the previous sections, longitudinal deformations along $\boldsymbol{k}$ affect the rapidity distribution and forward-backward asymmetry, while transverse components in the $\boldsymbol{i}$--$\boldsymbol{j}$ plane modify the azimuthal distribution (see eq. \eqref{eq:sin}). Projecting $\boldsymbol{\beta}$ onto the laboratory frame basis yields
\begin{subequations}
\begin{align}
\beta_x(t)&=\boldsymbol{\beta}\cdot\boldsymbol{i}=-\cos\chi\cos\delta\sin a+\sin\chi\left(\cos a\sin \omega t+\sin\delta \sin a \cos\omega t\right),\\
\beta_y(t)&=\boldsymbol{\beta}\cdot\boldsymbol{j}= \cos\chi\sin\delta      +\sin\chi\cos\delta\cos\omega t\,,\\
\beta_z(t)&=\boldsymbol{\beta}\cdot\boldsymbol{k}= \cos\chi\cos\delta\cos a+\sin\chi\left(\sin a\sin \omega t- \sin\delta \cos a \cos\omega t\right)\,.
\end{align}
\end{subequations}

Two experimental strategies can be pursued in the search for NC effects. One is to measure angular distributions as functions of sidereal time, as done by the CMS collaboration \cite{CMS:2024rcv} in their search for Lorentz-violating effects in top-quark pair production at the LHC, taking Earth's rotation into account. Alternatively, one can integrate over long data-taking periods and consider time-averaged observables. For the latter approach, the time-averaged components of $\boldsymbol{\beta}$,
\begin{equation}
\overline{\beta_i}=\frac{1}{T}\int_0^T\beta_i(t)\,\d t\,,
\end{equation}
are given by
\begin{subequations}
\begin{align}
\overline{\beta_x}&=-\cos\chi\cos\delta\sin a\,,\\
\overline{\beta_y}&=\cos\chi\sin \delta\,,\\
\overline{\beta_z}&=\cos\chi \cos\delta \cos a\,,
\end{align}
\end{subequations}
where the time-dependent sinusoidal terms average to zero.

Unlike scenarios in which NC effects enter at $\mathcal{O}(\Theta^2)$ (e.g., Refs.~\cite{Das:2011iq, Manohar:2014zca}), the linear dependence here allows a direct substitution of the averaged components of $\boldsymbol{\beta}$ into the cross sections and angular distributions.

For concreteness, we consider the ATLAS detector, located at $\delta \approx 46.236^\circ$, with a beam axis oriented approximately eastward ($a\approx 257.5^\circ$). \footnote{These values are inferred from the CERN map (\url{maps.cern.ch}).} In this case, the longitudinal component $\overline{\beta}_z$ is suppressed ($\cos a \approx -0.216$), while the transverse components remain sizable ($\sin a \approx -0.976$). Explicitly,
\begin{equation}
\overline{\boldsymbol \beta} = \cos\chi\left(0.675,0.722,-0.150\right).
\end{equation}
In fact, the maximum sensitivity to NC effects occurs when the deformation vector $\boldsymbol\beta$ is aligned with Earth's rotation axis ($\chi\to 0^\circ$ or $180^\circ$), whereas deformations in the equatorial plane lead to vanishing corrections in this setup.

In the following section we perform a comparison of our results for the NC distribution with experimental data from the ATLAS collaboration, accounting for Earth rotation effects.

\section{Comparison to ATLAS data}

In this section, we constrain the predictions of the NCSM using experimental data from the ATLAS experiment at the LHC, focusing on $Z(\mu\mu)$+jet events~\cite{ATLAS:2024xxl,atlas_collaboration_2024_11507450}. The analysis uses the full Run-2 dataset, corresponding to an integrated luminosity of $139~\mathrm{fb}^{-1}$ of $pp$ collisions at $\sqrt{s}=13~\mathrm{TeV}$. We use an unbinned, particle-level dataset obtained by ATLAS using machine-learning-based unfolding techniques, which enables the extraction of observables directly from reconstructed kinematics.

The fiducial phase space is defined by the following selections: two oppositely charged muons originating from the primary vertex, dressed with photons within $\Delta R < 0.5$ to account for final-state radiation. Each muon must satisfy $p_{t}^{\mu} > 25~\mathrm{GeV}$ and $|\eta^{\mu}| < 2.5$. The dimuon invariant mass is required to be within $81~\mathrm{GeV} < m_{\mu\mu} < 101~\mathrm{GeV}$, and the reconstructed $Z$ boson must have $p_{t}^{Z} > 200~\mathrm{GeV}$. This high-$p_t$ requirement is crucial, as sensitivity to NC effects increases with the boson's transverse momentum. Jets are reconstructed using the anti-$k_t$ algorithm~\cite{Cacciari:2008gp} with a radius parameter $R=0.4$, and are required to have $p_t > 5~\mathrm{GeV}$ and $|y| < 2.5$.

{The theoretical predictions entering the comparison with ATLAS data are evaluated within the same fiducial phase space as the experimental analysis, with the full set of ATLAS event-selection requirements and kinematic cuts implemented in our MC simulation.}

The ATLAS detector employs a right-handed coordinate system centered at the interaction point, with the $z$-axis along the beam pipe, the $x$-axis pointing toward the LHC ring center, and the $y$-axis directed upward. For our NC analysis, we adopt {two complementary strategies. First, we fix the orientation of $\boldsymbol \beta$ that maximizes sensitivity to NC contributions and derive bounds on the NC scale $\Lambda$. Second, we fix $\Lambda$ and investigate how the predicted signals depend on the angular orientation of $\boldsymbol \beta$. In what follows, we perform this analysis using two observables: the azimuthal distribution and the forward-backward asymmetry.}

\subsection{{Azimuthal distribution}}

\begin{figure}[ht]
\centerline{\includegraphics[width=.7\textwidth]{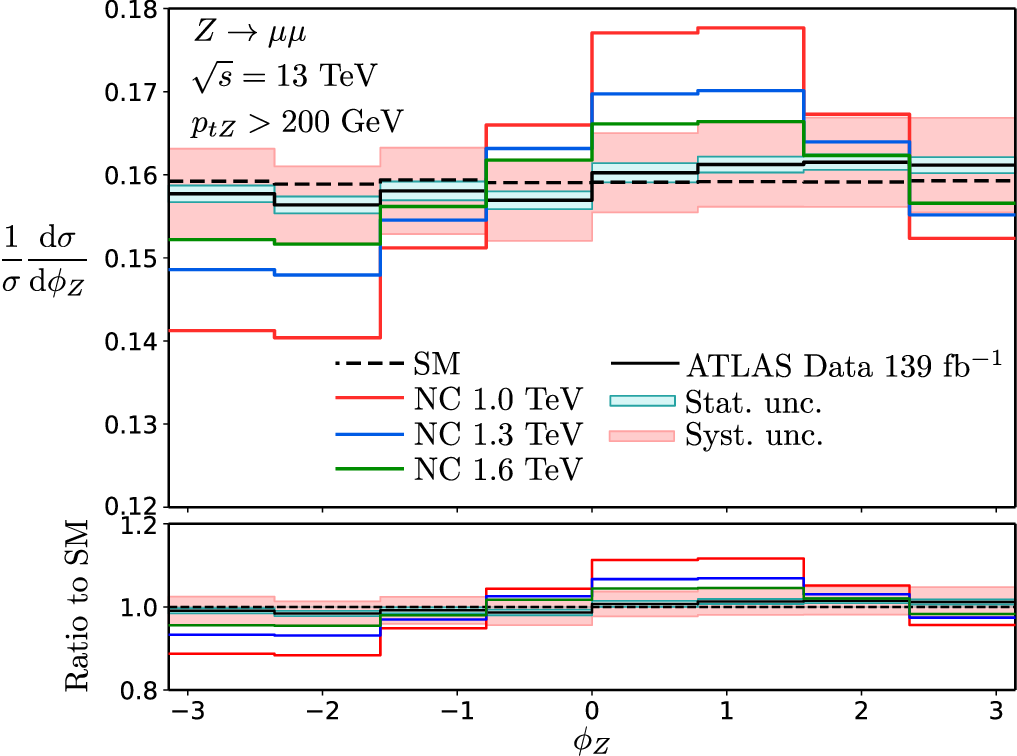}}
\vspace*{8pt}
\caption{Azimuthal distribution ($\phi_Z$) of the $Z$ boson obtained using ATLAS data, compared to predictions from the NCSM at different NC scales $\Lambda$. Statistical and systematic uncertainties are indicated. {For this plot we have set $\chi = 180^\circ$.}\label{fig6}}
\end{figure}

Figure~\ref{fig6} shows the azimuthal distribution of the reconstructed $Z$ boson, obtained from the $418\,014$ selected events. The data are compared to NCSM predictions for different NC scales $\Lambda$, {for a fixed orientation $\chi = 180^\circ$, corresponding to $\overline{\boldsymbol \beta} = \left(-0.675,-0.722,0.150\right)$}. Both statistical and systematic uncertainties are indicated; the latter account for detector effects, background contributions, the unfolding procedure, and uncertainties in the machine-learning model, among other sources.

Inspection of Fig.~\ref{fig6} reveals that bins 1, 2, 5, and 6 exhibit the greatest sensitivity to NC effects. For $\Lambda \simeq {1.6}~\mathrm{TeV}$, the predicted deviations in these bins are comparable to the experimental uncertainties, suggesting that current ATLAS data can exclude NC scales below $\sim {1.6}~\mathrm{TeV}$. The remaining bins show limited sensitivity.

\begin{figure}[ht]
\centerline{\includegraphics[width=.7\textwidth]{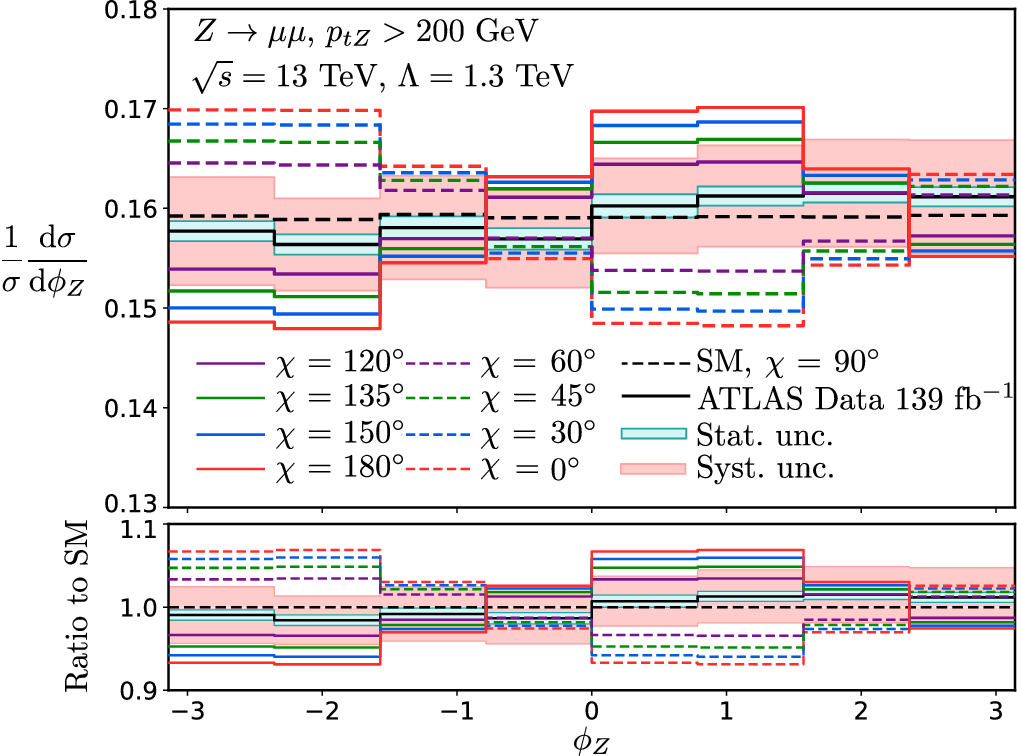}}
\vspace*{8pt}
\caption{{Azimuthal distribution ($\phi_Z$) of the $Z$ boson measured with ATLAS data, compared to NCSM predictions for different orientations of the NC deformation vector $\chi$. For this plot, we set $\Lambda=1.3$~TeV.}\label{fig6_chi}}
\end{figure}

{In Fig. \ref{fig6_chi}, we show the same distribution for a fixed NC scale $\Lambda=1.3$ TeV while varying the orientation angle $\chi$ of the NC deformation. This figure illustrates that Earth-rotation effects effectively reduce the amplitude of the sinusoidal structure in the NC azimuthal distribution by a factor of $\cos\chi$. For $\chi<90^\circ$, the phase of the oscillation is reversed relative to the case $\chi>90^\circ$. At this scale, orientations of $\boldsymbol{\beta}$ in the ranges $135^\circ\lesssim\chi\leq180^\circ$ and $0\leq\chi\lesssim 45^\circ$ exhibit strong deviations from the SM prediction, whereas deformations with $\boldsymbol\beta$ lying closer to the equatorial plane lead to reduced sensitivity.}

Overall, the ATLAS measurements are consistent with the SM prediction within uncertainties, providing no conclusive evidence for non-commutativity. However, an intriguing pattern is observed: for $\phi_Z < 0$, the data points tend to lie below the SM prediction, while for $\phi_Z > 0$ they tend to lie above it. This effect qualitatively resembles the sinusoidal modulation predicted in Eq.~(\ref{eq:sin}). {Moreover, the overall structure of the data with uncertainties appears to suggest a sinusoidal pattern compatible with an deformation angle $\chi>90^\circ$}, though it remains inconclusive given the current size of the systematic uncertainties. Future improvements in systematic control could clarify whether such a trend is physical.

\subsection{{Forward-backward Asymmetry}}

\begin{figure}[ht]
\centerline{\includegraphics[width=.9\textwidth]{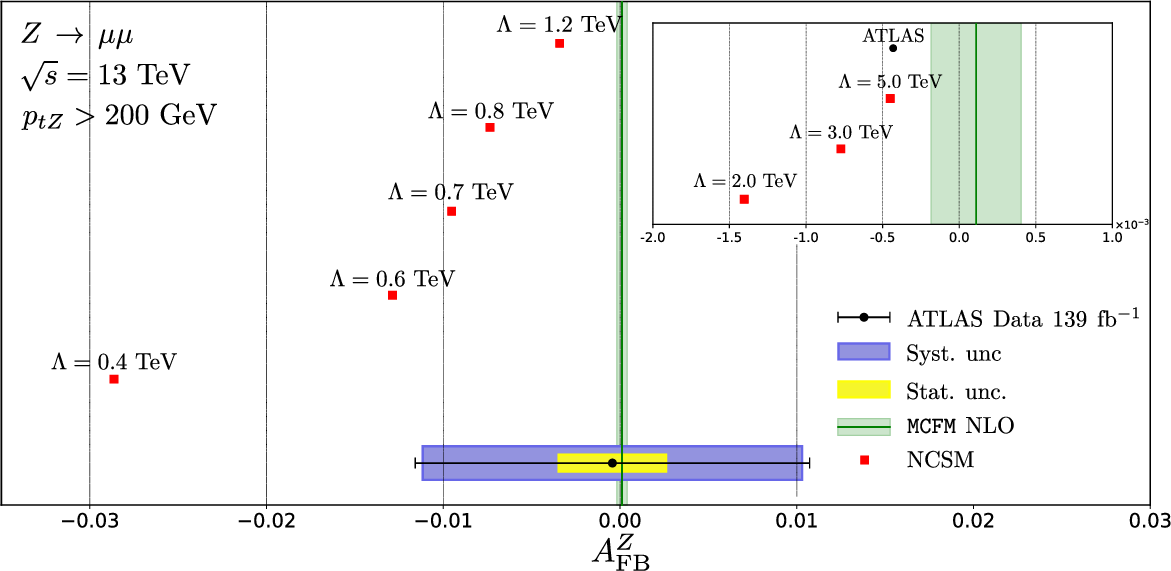}}
\vspace*{8pt}
\caption{Forward-backward asymmetry of the $Z$ boson obtained using ATLAS data, compared to NCSM predictions at various $\Lambda$ {and fixed $\chi=0^\circ$}, and with the SM prediction from \texttt{MCFM} at NLO. The inset shows a zoomed-in view to illustrate the constraint derived from the central value.\label{fig7}}
\end{figure}

Turning to the forward-backward asymmetry, from the ATLAS data we extract
\begin{equation}
A_{\mathrm{FB}}^{Z(\mathrm{ATLAS})} = -0.00043 \pm 0.0030 \, (\mathrm{stat}) \pm 0.0107 \, (\mathrm{syst}).
\end{equation}
This result is shown in Fig.~\ref{fig7}, compared to the NCSM predictions at different $\Lambda$ values and {fixed $\chi=0^\circ$, together with} the SM prediction obtained from \texttt{MCFM} at NLO (with Monte Carlo integration errors and scale uncertainties):
\begin{equation}
A_{\mathrm{FB}}^{Z(\texttt{MCFM})} = 0.00011 \pm 0.00029 \, (\mathrm{MC})^{+0.00004}_{-0.00003} \, (\mathrm{scale}).
\end{equation}

{Although the experimental uncertainties are sizable, the central value from the ATLAS measurement is compatible with the SM prediction. Including both statistical and systematic uncertainties yields a robust lower exclusion bound on the NC scale of $\Lambda \gtrsim {0.6}~\mathrm{TeV}$. If only the statistical uncertainty is retained, one obtains a more optimistic sensitivity estimate of $\Lambda \sim {1.2}~\mathrm{TeV}$, rather than a comparably robust exclusion limit. Likewise, if the central experimental value of $A_{\mathrm{FB}}$ is used directly as an idealized constraint (see inset of Fig.~\ref{fig7}), the implied sensitivity reach can extend to $\Lambda \sim {5}~\mathrm{TeV}$. Extrapolating to the future High-Luminosity LHC (HL-LHC) phase, where both statistical and systematic uncertainties are expected to improve, the projected sensitivity to NC scales could realistically reach $\mathcal{O}(10~\mathrm{TeV})$.}

{The relatively modest lower bound on $\Lambda$ obtained from $A_{\mathrm{FB}}^{Z}$ can be traced to the suppression of the beam-axis component, $\overline{\beta}_z=0.150$. This suppression originates from the orientation of the ATLAS beam axis, being close to the eastward direction, which reduces sensitivity to NC effects associated with the longitudinal component. In principle, stronger bounds could be obtained from detectors whose beam axes are oriented closer to the North--South direction, where sensitivity to NC effects along the beam axis is enhanced.}

\begin{figure}[ht]
\centerline{\includegraphics[width=.7\textwidth]{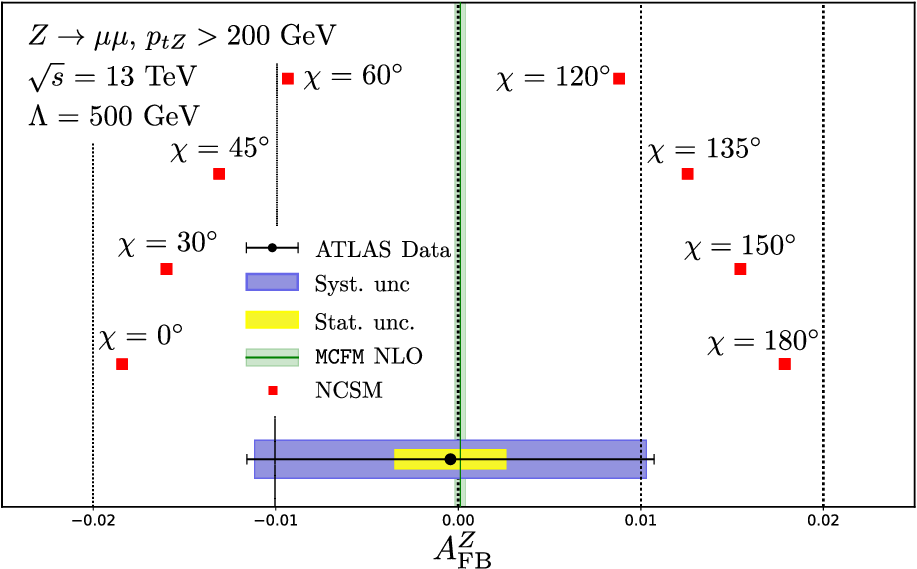}}
\vspace*{8pt}
\caption{Forward-backward asymmetry of the $Z$ boson obtained using ATLAS data, compared to NCSM predictions at various $\Lambda$ and the SM prediction from \texttt{MCFM} at NLO. The inset shows a zoomed-in view to illustrate the constraint derived from the central value. {For this plot we have set $\chi = 0^\circ$.}\label{fig7_chi}}
\end{figure}

{We also show in Fig.~\ref{fig7_chi} the behavior of $A^Z_{\mathrm{FB}}$ for different orientations of the NC deformation angle $\chi$, keeping the NC scale fixed at $\Lambda=500$~GeV. In particular, $A^Z_{\mathrm{FB}}$ changes sign as $\chi$ varies across the equatorial configuration near $\chi \sim 90^\circ$. As noted before, the largest deviations from the SM prediction occur for orientations near $\chi=0^\circ$ and $180^\circ$, while configurations around $\chi\approx 90^\circ$ yield strongly suppressed asymmetries. Interestingly, several orientations at $\Lambda=500$~GeV remain compatible with current ATLAS uncertainties, emphasizing that collider constraints on noncommutativity can depend sensitively not only on the scale $\Lambda$, but also on the geometric orientation of the underlying deformation.}

Overall, the azimuthal distribution and forward-backward asymmetry emerge as powerful and complementary observables for testing NC geometry at the LHC. While current ATLAS measurements are consistent with the SM, they already constrain the NC scale in the {TeV} domain, with {conservative exclusion limits of $\Lambda \gtrsim {0.6}~\mathrm{TeV}$ from $A_{\mathrm{FB}}$ and $\Lambda \gtrsim {1.6}~\mathrm{TeV}$ from the azimuthal distribution, while more optimistic assumptions suggest sensitivity reaches up to $\Lambda \sim {5}~\mathrm{TeV}$.} These findings highlight the potential of precision $Z$+jet measurements as indirect probes of new physics. Looking ahead, the HL-LHC will significantly improve both the statistical and systematic precision of such measurements, potentially extending the sensitivity reach for NC geometry into the $\mathcal{O}(10~\mathrm{TeV})$ regime.

\subsection{{Comparison with bounds from other processes}}

Having obtained lower bounds on the NC scale $\Lambda$ from the azimuthal distribution and the forward-backward asymmetry in $Z+\text{jet}$ production, we now compare our findings with those reported in previous studies. Table~\ref{tab:Lambda_bounds_full} summarizes the existing lower bounds on $\Lambda$ from various collider processes and experiments. The bounds derived in the present work are listed in the last two rows.

\begin{table}[ht]
\tbl{Representative existing lower bounds (and projected sensitivities, where indicated) on the NC scale $\Lambda$ from collider processes. The bounds derived in the present analysis are listed in the final two rows.}
{
\begin{tabular}{@{}llcl@{}}
\toprule
Process  & Experiment & Bound (TeV) & Reference \\
\colrule
$e^+e^- \to \gamma\gamma$ & LEP (OPAL) & $0.141$ & \cite{OPAL:2003eoc} \\
$e^-e^-\to e^-e^-$ & Linear Colliders (Projected) & $1.7$ & \cite{Hewett:2000zp} \\
$e^-\gamma \to e^-\gamma$ & NLC (projected) & $1.0$--$2.5$ & \cite{Mathews:2000we} \\
$\gamma\gamma\to \gamma\gamma$ & LHC (projected) & $1.64$ & \cite{Inan:2022aiw} \\
$pp\to Z\gamma$ & LHC (projected) & $1.0$ & \cite{Alboteanu:2006hh} \\
$pp \to X\gamma$ & LHC & $1.145$ & \cite{Bekli:2020unl} \\
$pp \to t\bar{t}$ & Tevatron (CDF+D\O) & $0.7$ & \cite{Fisli:2020vzt} \\
$pp \to Z(\mu\mu)+\text{jet}$ ($\phi$) & LHC (ATLAS)  & $1.6$ & This work \\
$pp \to Z(\mu\mu)+\text{jet}$ ($A_{\mathrm{FB}}$) & LHC (ATLAS) & $0.6$ & This work \\
\botrule
\end{tabular}
\label{tab:Lambda_bounds_full}}
\end{table}

The LEP bound from $e^+e^- \to \gamma\gamma$ ~\cite{OPAL:2003eoc} is notable as a direct constraint derived from a clean leptonic environment, whereas several of the other bounds, including those obtained here, arise from phenomenological analyses of hadron-collider observables. Our bounds, ranging from $\Lambda \gtrsim 0.6$ TeV to $\Lambda \gtrsim 1.6$ TeV depending on the observable, are competitive with and complementary to existing bounds in the literature.

\section{Conclusions}

In this work, we have conducted a comprehensive study of $W/Z$+jet production at the LHC within the framework of the NCSM. Starting from the SW map, we derived the leading-order squared amplitudes for all relevant partonic subprocesses, demonstrating that NC corrections emerge at $\mathcal{O}(\Theta)$, a distinctive feature that enhances the sensitivity of vector-boson+jet production to NC effects, in contrast to many other processes where leading corrections appear only at $\mathcal{O}(\Theta^2)$.

We performed a detailed phenomenological analysis using a Monte Carlo reweighting technique. Our results show that while the total production cross-section remains largely insensitive to space-space non-commutativity, key angular observables exhibit significant sensitivity. In particular, the forward-backward asymmetry in the rapidity distribution probes NC deformations along the beam axis, while the azimuthal distribution provides a powerful probe of anisotropies in the transverse plane. These two observables offer complementary handles on the structure of spacetime non-commutativity.

{An important aspect addressed in this work is the impact of Earth rotation on NC signals. Because the NC deformation is defined in a fixed celestial frame while collider observables are measured in the rotating laboratory frame, sidereal effects can modify the effective orientation of the deformation seen by the detector. In particular, depending on the location and orientation of the detector, Earth rotation can partially wash out otherwise sizable NC signals for unfavorable orientations.}

By comparing our theoretical predictions with unfolded, particle-level ATLAS Run-2 data for $Z(\mu\mu)$+jet events, we extracted robust constraints on the NC scale $\Lambda$. The measured azimuthal distribution and forward-backward asymmetry are consistent with the SM within current uncertainties, allowing us to set lower bounds of $\Lambda \gtrsim {0.6\text{--}1.6}~\mathrm{TeV}$ under conservative assumptions, with sensitivity extending up to $\Lambda \sim {5}~\mathrm{TeV}$ for more aggressive interpretations. {Accounting for Earth-rotation effects shows that these limits depend on the orientation of the NC deformation.} These results firmly establish LHC vector-boson+jet channels as powerful probes of noncommutative geometry.

Looking ahead, the high-luminosity phase of the LHC will deliver significantly reduced experimental uncertainties. This advancement, combined with future theoretical work including next-to-leading-order corrections within the NCSM and multivariate analyses, will push the sensitivity to NC scales well into the $\mathcal{O}(10~\mathrm{TeV})$ regime. Our study therefore highlights both the current constraints and the promising prospects for testing the fundamental structure of spacetime at the LHC.

\section*{Acknowledgments}

This work was supported by PRFU research project No. B00L02UN050120230003. The authors gratefully acknowledge financial support from the Algerian Ministry of Higher Education and Scientific Research and the Directorate General for Scientific Research and Technological Development (DGRSDT).

Part of the numerical computations presented in this work were performed using the HPC cluster at the University of Batna 2.

\appendix

\section{Feynman rules}\label{sec:Feyn}

\begin{figure}[ht]
\centerline{
\begin{minipage}{0.25\textwidth}
\centering
\includegraphics[width=.95\textwidth]{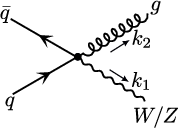}
\end{minipage}
\begin{minipage}{0.75\textwidth}
$$\begin{aligned}
\mathcal{V}_{q\to Wgq}&=\frac{eg_s}{4\sqrt{2}\sin\theta_w}\,V_{ij}\,t^a\,\mathcal{T}_{\mu\nu}(k_1,k_2)(1-\gamma_5)\\
\mathcal{V}_{q\to Zgq}&=\frac{eg_s}{2\sin2\theta_w}\,t^a\,\mathcal{T}_{\mu\nu}(k_1,k_2)(c_V-c_A\gamma_5)\\
\mathcal{T}_{\mu\nu}(k_1,k_2)&=\Theta_{\mu\nu}\left(\slashed{k}_1-\slashed{k}_2\right)+\gamma_\mu[\Theta(k_1-k_2)]_\nu\\
&\quad+\gamma_\nu[(k_1-k_2)\Theta]_\mu
\end{aligned}$$
\end{minipage}}
\centerline{
\begin{minipage}{0.25\textwidth}
\centering
\includegraphics[width=.7\textwidth]{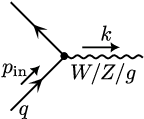}
\end{minipage}
\begin{minipage}{0.75\textwidth}
$$\begin{aligned}
\mathcal{V}_{q\to Wq}&=\frac{ie}{2\sqrt{2}\sin\theta_w}\,V_{ij}\,\Gamma_{\mu}(p_{\mathrm{in}},k)(1-\gamma_5)\\
\mathcal{V}_{q\to Zq}&=\frac{ie}{\sin2\theta_w}\,\Gamma_{\mu}(p_{\mathrm{in}},k)(c_V-c_A\gamma_5)\\
\mathcal{V}_{q\to gq}&=i\,g_s\,t^a\,\Gamma_{\mu}(p_{\mathrm{in}},k)\\
\Gamma_\mu(p_{\mathrm{in}},k)&=\gamma_\mu+\frac{i}{2}\left(p_{\mathrm{in}}\Theta \right)_\mu \slashed{k}
+\frac{i}{2}\gamma_\mu\left(k\Theta p_{\mathrm{in}}\right)
    +\frac{i}{2}\left(\Theta k\right)_\mu \slashed{p}_{\mathrm{in}}
\end{aligned}$$
\end{minipage}
}
\vspace*{8pt}
\caption{Relevant 4-point and 3-point vertices in the NCSM. The momentum $p_{\mathrm{in}}$ flows into the vertex, while $k$ and $k_i$ flow out.\protect\label{fig8}}
\end{figure}

Gauge invariance, and the consequent validity of Ward identities, must be maintained in the presence of NC modifications~\cite{Ohl:2004tn,Ohl:2004ke}. This requirement enforces the inclusion of a novel four-point interaction vertex~\cite{Alboteanu:2006hh}, absent in the SM, as shown in Fig.~\ref{fig1}. The relevant Feynman rules for this diagram, along with the deformed triple gauge vertices in the massless quark limit, are displayed in Fig.~\ref{fig8}~\cite{Calmet:2001na,Melic:2005fm,Melic:2005am}. Here, $t^a$ denotes the generators of $\mathrm{SU}(N)$ in the fundamental representation, $p_{\mathrm{in}}$ is the momentum of the incoming quark, and $k_i$ are the momenta of the outgoing vector bosons.

It is worth noting that our vertex conventions differ from those in Refs.~\cite{Calmet:2001na,Melic:2005fm,Melic:2005am} by relative signs. This is a pure convention choice, arising from our definition that all vector boson momenta ($k, k_i$) flow \emph{out} of the vertex.

For an \emph{external} incoming quark, the terms proportional to $\slashed{p}_{\mathrm{in}}$ in the three-point vertices vanish upon acting on the external spinor ($\slashed{p}_{\mathrm{in}} u(p_{\mathrm{in}}) = 0$ for massless on-shell fermions). The same holds for external outgoing quarks. Moreover, terms proportional to $k\,\Theta\, p$ contribute only at $\mathcal{O}(\Theta^2)$.

To see this, observe that such terms modify the SM amplitude schematically as
\begin{equation}
\mathcal{M}_t+\mathcal{M}_u\rightarrow \mathcal{M}_t\left[1+i(k\Theta p)\right]+ \mathcal{M}_u\left[1+i(k'\Theta p')\right],
\end{equation}
where $\mathcal{M}_t$ and $\mathcal{M}_u$ are the SM $t$- and $u$-channel amplitudes. At $\mathcal{O}(\Theta)$, this deformation generates an interference term proportional to $\mathcal{M}_t \mathcal{M}_u^* - \mathcal{M}_u \mathcal{M}_t^*$, which vanishes since the SM amplitudes are relatively real at tree level.

Furthermore, in the four-point vertex, only the term proportional to $\Theta_{\mu\nu}$ contributes at $\mathcal{O}(\Theta)$, as the remaining two tensor structures cancel. Consequently, the effective vertex functions relevant for our leading-order analysis simplify to
\begin{subequations}
\begin{align}
\Gamma_\mu(p_{\mathrm{in/out}},k)&=\gamma_\mu+\frac{i}{2}\left(p_{\mathrm{in/out}}\Theta \right)_\mu \slashed{k}\,,\\
\mathcal{T}_{\mu\nu}(k_1,k_2)&=\Theta_{\mu\nu}\left(\slashed{k}_1-\slashed{k}_2\right),
\end{align}
\end{subequations}
where $p_{\mathrm{in/out}}$ denotes the momentum of an external quark (incoming or outgoing). For external antiquarks, the momentum is assigned a minus sign relative to the fermion flow direction; for example, an incoming antiquark carries momentum $-p_{\mathrm{in}}$. 

\section{Squared decay amplitudes} \label{sec:decay}

In this appendix, we explicitly demonstrate that the squared amplitudes for the leptonic decays $Z \to {\mu^-\mu^+}$ and $W \to e\nu$ receive corrections from NC geometry only at $\mathcal{O}(\Theta^2)$. This observation justifies neglecting such effects in the leading-order phenomenological analysis presented in the main text, which retains terms only up to $\mathcal{O}(\Theta)$.

Consider first the process $Z(k)\to {\mu}^-(p)\,{\mu}^+(p')$. Stripping off overall constant factors, the decay amplitude can be written as
\begin{subequations}
\begin{align}
i\widetilde{\mathcal{M}}&=\bar{u}(p)\,\Gamma_\mu\left[p,-k\right](c_V-c_A\gamma_5)\,v(p')\,\epsilon^{\mu}(k)=i\widetilde{\mathcal{M}}_{\mathrm{SM}}+i\widetilde{\mathcal{A}}_{\mathrm{NC}}\,,\\
i\widetilde{\mathcal{M}}_{\mathrm{SM}}&=\bar{u}(p)\,\gamma_\mu\,(c_V-c_A\gamma_5)\,v(p')\,\epsilon^{\mu}(k)\,,\\
i\widetilde{\mathcal{A}}_{\mathrm{NC}}&=-\frac{i}{2}\,(p\,\Theta)_\mu\,\bar{u}(p)\,\slashed{k}\,(c_V-c_A\gamma_5)\,v(p')\,\epsilon^{\mu}(k)\,.
\end{align}
\end{subequations}
The squared amplitude takes the form
\begin{subequations}
\begin{align}
|\widetilde{\mathcal{M}}|^2&=|\widetilde{\mathcal{M}}_{\mathrm{SM}}|^2+|\widetilde{\mathcal{M}}_{\mathrm{NC}}|^2\,,\\
|\widetilde{\mathcal{M}}_{\mathrm{SM}}|^2&=\bar{u}(p)\,\gamma_\mu\,(c_V-c_A\gamma_5)\,v(p')\,\bar{v}(p')\,\gamma_\nu\,(c_V-c_A\gamma_5)\,u(p)\,\epsilon^{\mu}(k)\,\epsilon^{\nu*}(k)\,,\\
|\widetilde{\mathcal{M}}_{\mathrm{NC}}|^2&=\left[-i\widetilde{\mathcal{M}}^*_{\mathrm{SM}}\,i\widetilde{\mathcal{A}}_{\mathrm{NC}}+\mathrm{c.c.}\right]+|\mathcal{A}^{}_{\mathrm{NC}}|^2\,,
\end{align}
\end{subequations}
where c.c.~denotes complex conjugation.

After spin summation and polarization averaging, the SM contribution yields
\begin{align}
\overline{|\widetilde{\mathcal{M}}_{\mathrm{SM}}|^2}&=\frac{1}{3}\,\mathrm{tr}\left[(\slashed{p}+m_e)\gamma_\mu(c_V-c_A\gamma_5)(\slashed{p}'-m_e)\gamma_\nu(c_V-c_A\gamma_5)\right]\left[-g^{\mu\nu}+\frac{k^\mu k^\nu}{M_Z^2}\right]\notag\\
&=\frac{4}{3} \left[2\left(c_V^2-2\,c_A^2\right) m_e^2+\left(c_A^2+c_V^2\right) m_Z^2\right].
\end{align}
The interference between the SM and NC contributions vanishes. Explicitly,
\begin{align}
\left[\overline{\widetilde{\mathcal{M}}^*_{\mathrm{SM}}\,\widetilde{\mathcal{A}}_{\mathrm{NC}}}+\mathrm{c.c.}\right]&=-\frac{i}{6}\,(p\,\Theta)_\nu\,\mathrm{tr}\left[(\slashed{p}+m_e)\slashed{k}(c_V-c_A\gamma_5)(\slashed{p}'-m_e) \gamma_\mu(c_V-c_A\gamma_5)\right]\notag\\
&\quad\times\left[-g^{\mu\nu}+\frac{k^\mu k^\nu}{M_Z^2}\right]+\text{c.c.}\notag\\
&=-\frac{8}{3}\,c_V\,c_A\,\epsilon^{\nu kpp'}\,(p\,\Theta)_\nu=0\,,
\end{align}
where the last equality follows from the antisymmetry of the Levi-Civita tensor together with $k=p+p'$.

We are thus left only with the quadratic NC correction,
\begin{align}
\overline{|\widetilde{\mathcal{M}}_{\mathrm{NC}}|^2}&=\overline{|\mathcal{A}^{}_{\mathrm{NC}}|^2}=\frac{1}{12}\,(p\,\Theta)_\mu(p\,\Theta)_\nu\notag\\
&\times\mathrm{tr}\left[(\slashed{p}+m_e)\slashed{k}(c_V-c_A\gamma_5)(\slashed{p}'-m_e)\slashed{k}(c_V-c_A\gamma_5)\right]\left[-g^{\mu\nu}+\frac{k^\mu k^\nu}{M_Z^2}\right]\notag\\
&=\frac{2}{3}\,c_A^2\,m_e^2\left[(p\,\Theta\,p')^2-m_Z^2\,(p\,\Theta)^2\right].
\end{align}
which is explicitly of order $\Theta^2$. 

A completely analogous calculation for $W(k)\to e(p)\,\nu(p')$ yields, up to overall couplings,
\begin{equation}
\overline{|\widetilde{\mathcal{M}}|^2}=
\frac{4 \left(2 M_W^4-m_e^2 M_W^2-m_e^4\right)}{3 M_W^2}
+\frac{m_e^2(M_W^2-m_e^2)}{3M_W^2}\left[(p\,\Theta p')^2-M_W^2\,(p\,\Theta)^2\right].
\end{equation}

In conclusion, both $Z$ and $W$ leptonic decays receive corrections from NC geometry only at $\mathcal{O}(\Theta^2)$. As a result, leptonic decay widths are far less sensitive probes of NC geometry than production cross-sections or angular observables, which exhibit leading $\mathcal{O}(\Theta)$ effects and therefore provide the primary sensitivity discussed in the main text.

\section{Squared production amplitudes} \label{sec:Amps}

In this appendix, we present the analytic calculation of the Born-level squared amplitudes for the two relevant partonic channels within the NCSM
\begin{align*}
(\delta_g):\quad & q\bar{q} \to V + g\,, \\
(\delta_q):\quad & qg \to V + q\,,
\end{align*}
where $V$ denotes either a $Z$ or $W$ boson.

\subsection{Gluon channel}

We begin with the gluon channel, corresponding to $q(p_1)\bar{q}'(p_2)\to W/Z(k_1)g(k_2)$, as depicted in Fig.~\ref{fig1}. The tree-level amplitude, stripped of coupling constants, is
\begin{align}
i\mathcal{M}^{\delta_g}&=\bar{v}(p_2) \Gamma_\mu\left[-p_2,k_1\right] (c_V-c_A\gamma_5)\frac{i(\slashed{k}_1-\slashed{p}_2)}{(k_1-p_2)^2}\Gamma_\nu\left[p_1,k_2\right] u(p_1) \epsilon^{\mu *}(k_1)\epsilon^{\nu *}(k_2)\notag\\
&+\bar{v}(p_2)\Gamma_\nu\left[-p_2,k_2\right]\frac{i(\slashed{k}_2-\slashed{p}_2)}{(k_2-p_2)^2}\Gamma_\mu\left[p_1,k_1\right] (c_V-c_A\gamma_5) u(p_1) \epsilon^{\mu *}(k_1)\epsilon^{\nu *}(k_2)\notag\\
&-\frac{1}{2}\Theta_{\mu\nu}\bar{v}(p_2)(\slashed{k}_1-\slashed{k}_2)(c_V-c_A\gamma_5)u(p_1)\epsilon^{\mu *}(k_1)\epsilon^{\nu *}(k_2)\notag\\
&=i\mathcal{M}^{\delta_g}_{\mathrm{SM}}+i\mathcal{A}^{\delta_g}_{\mathrm{NC}}\,,
\end{align}
where $c_V$ and $c_A$ are the vector and axial couplings (for $W$ production, $c_V = c_A = 1$). The SM contribution is obtained by the replacement $\Gamma_\mu \to \gamma_\mu$.

Using the properties of the Dirac algebra, the NC correction can be simplified to $\mathcal{O}(\Theta)$ as
\begin{align}
i\mathcal{A}^{\delta_g}_{\mathrm{NC}}&=-\frac{1}{2}(\Theta [p_1+p_2])_\nu\,\bar{v}(p_2)\gamma_\mu (c_V-c_A\gamma_5)u(p_1)\left[\epsilon^{\mu*}(k_1)\epsilon^{\nu *}(k_2)+\epsilon^{\mu*}(k_2)\epsilon^{\nu *}(k_1)\right]\notag\\
&-\Theta_{\mu\nu}\bar{v}(p_2)\slashed{k}_1(c_V-c_A\gamma_5)u(p_1)\epsilon^{\mu *}(k_1)\epsilon^{\nu *}(k_2).
\end{align}
The squared matrix element then reads
\begin{equation}
|\mathcal{M}^{\delta_g}|^2=|\mathcal{M}^{\delta_g}_{\mathrm{SM}}|^2+|\mathcal{M}^{\delta_g}_{\mathrm{NC}}|^2+\mathcal{O}(\Theta^2),
\end{equation}
with
\begin{equation}
|\mathcal{M}^{\delta_g}_{\mathrm{NC}}|^2=-i\mathcal{M}_{\mathrm{SM}}^{\delta_g*}\,i\mathcal{A}^{\delta_g}_{\mathrm{NC}}+\mathrm{c.c.}\,.
\end{equation}
The explicit SM conjugate amplitude is
\begin{align}
-i\mathcal{M}_{\mathrm{SM}}^{\delta_g*}&=-i\bar{u}(p_1) \gamma_\alpha \frac{\slashed{k}_1-\slashed{p}_2}{(k_1-p_2)^2}\gamma_\beta (c_V-c_A\gamma_5) v(p_2) \epsilon^{\beta}(k_1)\epsilon^{\alpha}(k_2)\notag\\
&-i\bar{u}(p_1) \gamma_\beta (c_V-c_A\gamma_5) \frac{\slashed{k}_2-\slashed{p}_2}{(k_2-p_2)^2}\gamma_\alpha v(p_2) \epsilon^{\beta}(k_1)\epsilon^{\alpha}(k_2)\,.
\end{align}

After spin and polarization summation, the interference terms can be organized as
\begin{align*}
|\mathcal{M}^{\delta_g}_{\mathrm{NC}}|^2&=\frac{i}{2}(\Theta [p_1+p_2])_\nu\left[\left(g^{\mu\beta}-\frac{k_1^\mu k_1^\beta}{M_V^2}\right)g^{\nu\alpha}+\mu\leftrightarrow \nu\right]\\
&\quad\times\bigg\{\frac{1}{u}\mathrm{tr}\left[\slashed{p}_1\gamma_\alpha(\slashed{k}_1-\slashed{p}_2)\gamma_\beta \slashed{p}_2\gamma_\mu(c_V-c_A\gamma_5)^2\right]\\
&\qquad+\frac{1}{t}\mathrm{tr}\left[\slashed{p}_1\gamma_\beta(\slashed{k}_2-\slashed{p}_2)\gamma_\alpha \slashed{p}_2\gamma_\mu(c_V-c_A\gamma_5)^2\right]\bigg\}+\mathrm{c.c.}\\
&\quad+\frac{i}{u}\Theta_{\mu\nu}\left[g^{\mu\beta}-\frac{k_1^\mu k_1^\beta}{M_V^2}\right]g^{\nu\alpha}\mathrm{tr}\left[\slashed{p}_1 \gamma_\alpha (\slashed{k}_1-\slashed{p}_2)\gamma_\beta  \slashed{p}_2 \slashed{k}_1(c_V-c_A\gamma_5)^2\right]+\mathrm{c.c.}\\
&\quad+\frac{i}{t}\Theta_{\mu\nu} \left[g^{\mu\beta}-\frac{k_1^\mu k_1^\beta}{M_V^2}\right]g^{\nu\alpha}\mathrm{tr}\left[\slashed{p}_1\gamma_\beta(\slashed{k}_2-\slashed{p}_2)\gamma_\alpha \slashed{p}_2
\slashed{k}_1(c_V-c_A\gamma_5)^2\right]+\mathrm{c.c.}
\end{align*}
where the explicit Dirac traces are evaluated using \texttt{FeynCalc} \cite{Shtabovenko:2020gxv}. Including the appropriate prefactors, the final expressions are given in Eq.~\eqref{eq:deltag}.  

For numerical evaluation, the relevant chiral asymmetry and helicity-sum couplings for the $Z$ boson are collected in Table~\ref{tab:zcouplings}.
\begin{table}[ht]
\tbl{Chiral asymmetry and helicity-sum couplings for the $Z$ boson.}
{\begin{tabular}{@{}lcc@{}} \toprule
Quark Type & $2c_V c_A$ & $c_V^2 + c_A^2$ \\
\colrule
Up-type ($u,\bar{u},c,\bar{c},t,\bar{t}$) & 0.204 & 0.291 \\
Down-type ($d,\bar{d},s,\bar{s},b,\bar{b}$) & 0.352 & 0.374 \\
\botrule
\end{tabular}\label{tab:zcouplings}}
\end{table}

\subsection{Quark channel}

For the $q(p_1)g(p_2) \to V(k_1)q'(k_2)$ process, where $V$ denotes either a $W$ or $Z$ boson, the amplitude can be written (stripped of coupling constants) as
\begin{align}
i\mathcal{M}^{\delta_q}&=\bar{u}(k_2)\Gamma_\mu\left[k_2,k_1\right](c_V-c_A\gamma_5)\frac{i(\slashed{k}_1+\slashed{k}_2)}{(k_1+k_2)^2} \Gamma_\nu\left[p_1,-p_2\right]u(p_1)\epsilon^{\mu *}(k_1)\epsilon^{\nu }(p_2)\notag\\
&+\bar{u}(k_2)\Gamma_\nu\left[k_2,-p_2\right]\frac{i(\slashed{k}_2-\slashed{p}_2)}{(k_2-p_2)^2} \Gamma_\mu\left[p_1,k_1\right](c_V-c_A\gamma_5)u(p_1)\epsilon^{\mu*}(k_1)\epsilon^{\nu}(p_2)\notag\\
&-\frac{1}{2}\Theta_{\mu\nu}\bar{u}(k_2)(\slashed{k}_1+\slashed{p}_2)(c_V-c_A\gamma_5)u(p_1)\epsilon^{\mu *}(k_1)\epsilon^{\nu }(p_2)\,.\label{eq:C}
\end{align}
Proceeding analogously to the gluon channel, the amplitude can be reorganized as
\begin{align}
i\mathcal{M}^{\delta_q}&=i\mathcal{M}^{\delta_q}_{\mathrm{SM}}+i\mathcal{A}^{\delta_q}_{\mathrm{NC}}\\
i\mathcal{A}^{\delta_q}_{\mathrm{NC}}&=-\frac{1}{2}(\Theta [p_1-k_2])_\nu\,\bar{u}(k_2)\gamma_\mu (c_V-c_A\gamma_5)u(p_1)\left[\epsilon^{\mu*}(k_1)\epsilon^{\nu}(p_2)+\epsilon^{\nu*}(k_1)\epsilon^{\mu}(p_2)\right]\notag\\
&-\Theta_{\mu\nu}\bar{u}(k_2)\slashed{k}_1(c_V-c_A\gamma_5)u(p_1)\epsilon^{\mu *}(k_1)\epsilon^{\nu}(p_2)\,,
\end{align}
with the SM contribution obtained by setting $\Gamma_\mu \to \gamma_\mu$ and $\Theta \to 0$ in Eq.~\eqref{eq:C}.

Squaring the amplitude gives
\begin{equation}
|\mathcal{M}^{\delta_q}|^2=|\mathcal{M}^{\delta_q}_{\mathrm{SM}}|^2+|\mathcal{M}^{\delta_q}_{\mathrm{NC}}|^2+\mathcal{O}(\Theta^2)\,,
\end{equation}
where the NC contribution arises from the interference term with the SM
\begin{align*}
|\mathcal{M}^{\delta_q}_{\mathrm{NC}}|^2&=-i\mathcal{M}^{\delta_q\,*}_{\mathrm{SM}}\,i\mathcal{A}^{\delta_q}_{\mathrm{NC}}+\mathrm{c.c.}\\
-i\mathcal{M}^{\delta_q\,*}_{\mathrm{SM}}&=-i\bar{u}(p_1) \gamma_\alpha\frac{\slashed{k}_1+\slashed{k}_2}{(k_1+k_2)^2}\gamma_\beta(c_V-c_A\gamma_5) u(k_2) \epsilon^{\beta}(k_1)\epsilon^{\alpha*}(p_2)\\
&-i\bar{u}(p_1) \gamma_\beta (c_V-c_A\gamma_5)\frac{\slashed{k}_2-\slashed{p}_2}{(k_2-p_2)^2}\gamma_\alpha u(k_2) \epsilon^{\beta}(k_1)\epsilon^{\alpha*}(p_2)\,.
\end{align*}

Performing the spin and polarization summation, the interference terms can be written as
\begin{align}
|\mathcal{M}^{\delta_q}_{\mathrm{NC}}|^2&=\frac{i}{2}(\Theta [p_1-k_2])_\nu \left[\left(g^{\mu\beta}-\frac{k_1^\mu k_1^\beta}{M_V^2}\right)g^{\nu\alpha}+\mu\leftrightarrow\nu\right]\notag\\
&\times\bigg\{\frac{1}{s}\mathrm{tr}\left[\slashed{p}_1\gamma_\alpha(\slashed{k}_1+\slashed{k}_2)\gamma_\beta \slashed{k}_2\gamma_\mu(c_V-c_A\gamma_5)^2\right]\notag\\
&+\frac{1}{t}\mathrm{tr}\left[\slashed{p}_1\gamma_\beta(\slashed{k}_2-\slashed{p}_2)\gamma_\alpha \slashed{k}_2\gamma_\mu(c_V-c_A\gamma_5)^2\right]\bigg\}+\mathrm{c.c.}\notag\\
&+\frac{i}{s}\Theta_{\mu\nu}\left[g^{\mu\beta}-\frac{k_1^\mu k_1^\beta}{M_V^2}\right]g^{\nu\alpha}\mathrm{tr}\left[\slashed{p}_1 \gamma_\alpha (\slashed{k}_1+\slashed{k}_2)\gamma_\beta \slashed{k}_2 \slashed{k}_1(c_V-c_A\gamma_5)^2\right] +\mathrm{c.c.} \notag\\
&+\frac{i}{t}\Theta_{\mu\nu} \left[g^{\mu\beta}-\frac{k_1^\mu k_1^\beta}{M_V^2}\right]g^{\nu\alpha}\mathrm{tr}\left[\slashed{p}_1\gamma_\beta(\slashed{k}_2-\slashed{p}_2)\gamma_\alpha \slashed{k}_2\slashed{k}_1(c_V-c_A\gamma_5)^2\right]+\mathrm{c.c.}
\end{align}
After evaluating the relevant Dirac traces using \texttt{FeynCalc}~\cite{Shtabovenko:2020gxv}, we obtain the result shown in Eq.~\eqref{eq:deltaq}.

For the crossed process $\bar{q}(p_2)g(k_2) \to W(k_1)\bar{q}(p_1)$, one obtains the same expression as for the quark channel.

\bibliographystyle{ws-mpla}
\bibliography{ref}

\end{document}